\RequirePackage{fix-cm}
\documentclass{svjour3}                     
\smartqed  
\usepackage[normalem]{ulem}
\usepackage{graphicx}
 \usepackage{subfigure}
\usepackage{amsmath,bm}
\usepackage{amsfonts}
\usepackage{amssymb}
\usepackage[sort,compress,numbers]{natbib}
\usepackage{enumitem}
\usepackage{textgreek}
\usepackage{appendix}
\usepackage[top=2.54cm, bottom=2.54cm, left=2.54cm, right=2.54cm]{geometry}
\usepackage{color}
\definecolor{cream}{RGB}{222,217,201}

\begin{document}

\title{Cell-level modelling of homeostasis in confined epithelial monolayers}

\author{Chaithanya K. V. S. \and Jan Rozman \and   Andrej Ko{\v s}mrlj   \and   Rastko Sknepnek }

\institute{Chaithanya K. V. S. \at
              School of Life Sciences, University of Dundee, Dundee DD1 5EH, United Kingdom.\\
              School of Science and Engineering, University of Dundee, Dundee DD1 4HN, United Kingdom. \\
              \email{chaithanyakvs@gmail.com}           
           \and
           Jan Rozman \at
              Rudolf Peierls Centre for Theoretical Physics, University of Oxford, \\ Oxford OX1 3PU, United Kingdom.
          \and
          Andrej Ko{\v s}mrlj \at
          Department of Mechanical and Aerospace Engineering, Princeton University, Princeton, New Jersey 08544, USA. \\
          Princeton Materials Institute, Princeton University, Princeton, New Jersey 08544, USA.   
          \and
          Rastko Sknepnek \at
          School of Life Sciences, University of Dundee, Dundee DD1 5EH, United Kingdom.\\
          School of Science and Engineering, University of Dundee, Dundee DD1 4HN, United Kingdom. \\
          \email{r.sknepnek@dundee.ac.uk} 
}

\date{Received: date / Accepted: date}

\maketitle

\begin{abstract}
Tissue homeostasis, the biological process of maintaining a steady state in tissue via control of cell proliferation and death, is essential for the development, growth,  maintenance, and proper function of living organisms. Disruptions to this process can lead to serious diseases and even death. In this study, we use the vertex model for the cell-level description of tissue mechanics to investigate the impact of the tissue environment and local mechanical properties of cells on homeostasis in confined epithelial tissues. We find a dynamic steady state,  where the balance between cell divisions and removals sustains homeostasis, and characterise the homeostatic state in terms of cell count, tissue area, homeostatic pressure, and the cells' neighbour count distribution. This work, therefore, sheds light on the mechanisms underlying tissue homeostasis and highlights the importance of mechanics in its control. 
\end{abstract}

\section{Introduction}
\label{intro}

Cell proliferation, the process by which cells grow and multiply through division, is essential for various biological functions such as tissue development, growth, and maintenance~\citep{alberts2017molecular}. For example, during the early stages of embryonic development cells rapidly proliferate, differentiate, and position themselves to lay out the body plan for the development of a new organism~\citep{wolpert2015principles}. Throughout adult life, maintaining tissue homeostasis involves a balance between cell proliferation and cell death. This is essential for tissue upkeep, repair, and regeneration in response to injury~\citep{landen2016transition,rodrigues2019wound, burclaff2020proliferation}. Disruptions of this balance can lead to serious diseases such as cancer~\citep{weinberg2013biology}, atherosclerosis~\citep{libby2021changing}, and rheumatoid arthritis~\citep{firestein2003evolving}. Therefore, a key question is how tissues maintain the balance between cell division and cell death to ensure homeostasis.

Tissue homeostasis relies on the delicate balance between cell proliferation and cell death, which are regulated not only by biochemical factors but also by mechanical cues \citep{duronio2013signaling,elmore2007apoptosis}. Cells experience forces by the surrounding tissues and extracellular matrix \citep{du2023tuning}. These mechanical forces affect cellular and intracellular biochemical signals that influence critical biological processes such as cell adhesion, migration, differentiation, and growth~\citep{vogel2006local}. A well-known example is contact inhibition, a process that halts cell division in dense environments to prevent tissue overcrowding and maintain integrity~\citep{stoker1967density,puliafito2012collective,lange2024minimal}. However, in many cancers, control of cell growth is disrupted, leading to a complex, heterogeneous mixture of actively dividing and quiescent cells, along with the necrotic tissue~\citep{gallaher2019impact,manzo2020defined}. In addition, the mechanical forces exerted by the microenvironment strongly impact cancerous growth by regulating the stresses imposed on a tumour, highlighting the critical role of mechanics in tumour progression~\citep{mierke2014fundamental}.  Despite its central importance, our understanding of the mechanical processes that control tissue homeostasis remains limited.

Furthermore, understanding the mechanical regulation of cell proliferation is important for identifying physical mechanisms that underlie the development of higher organisms. For example, in amniotes such as birds and reptiles, embryos before gastrulation~\citep{wolpert2015principles} (i.e. the developmental process in which an embryo transforms into a multilayered three-dimensional structure) are a flat disk of epithelial cells consisting of two main tissue types, the epiblast or embryonic tissue in the centre, and the extra-embryonic tissue encircling it~\citep{najera2020cellular}. Cell divisions and ingressions (i.e. removal of the cells into the region below the epiblast) occur throughout the epiblast, effectively maintaining its integrity during gastrulation~ \citep{najera2020cellular,najera2023evolution,asai2023coupling}. The extra-embryonic tissue, on the other hand, provides mechanical tension to the epiblast~\citep{downie1976mechanism}, which is essential for the proper execution of gastrulation. It also serves critical roles in nutrient transport, waste elimination, and providing protective barriers \citep{Treffkorn2022review}.    

In this paper, we explore how the tissue environment and cell mechanical properties contribute to establishing and maintaining homeostasis in confined planar epithelia. We start by briefly reviewing various approaches used to model the mechanics of epithelial cells, including particle-based models~\citep{szabo2006phase,sepulveda2013collective,matoz2017cell}, phase-field methods~\citep{marth2016collective, mueller2019emergence, moure2021phase, mueller2021phase}, cellular Potts models~\citep{glazier1993simulation, hirashima2017cellular}, Voronoi models~\citep{bi2016motility, barton2017active}, and, notably, vertex models~\citep{honda1978description,farhadifar2007influence,fletcher2013implementing,fletcher2014vertex}.

Particle-based models represent individual cells as discrete objects (typically discs or spheres) that interact with their neighbours via short-range potentials, as described e.g. in \cite{matoz2017cell,schnyder2017collective,kaiyrbekov2023migration}. While particle-based models can provide valuable insights into phenomena such as collective cell migration, they cannot easily account for the deformations and the resulting mechanical responses of cells. Additionally, since those models do not explicitly include cell-cell junctions, which play a key role in force transmission, defining precise interaction potentials is not straightforward.

Phase-field methods model cells using continuous density fields to describe cell shapes and the interfaces between them \cite{camley2014polarity,palmieri2015multiple,zhang2020active,monfared2023mechanical}. These methods can easily handle processes such as cell division, cell migration, and changes in cell shape. However, they are computationally expensive because each cell is described by a partial differential equation for its phase field, limiting accessible system sizes and times.

Cellular Potts models represent cells on a lattice, with each cell occupying multiple lattice sites \cite{graner1992simulation,chiang2016glass}. In this approach, it is straightforward to simulate processes such as cell sorting driven by differential adhesion~\cite{graner1992simulation,durand2021large} and tissue organization \cite{thuroff2019bridging}. It was also recently used to model cell competition \cite{carpenter2024physical}. While very powerful, the discrete nature of the cellular Potts model can, however, lead to artefacts, e.g. by limiting its ability to represent continuous mechanical properties and precise cellular deformations accurately.

In self-propelled Voronoi models the epithelial tissue is represented as a Voronoi tessellation of the plane. Cells are represented by particles that act as seed points of the Voronoi diagram. Each cell is, thus, a tile of the corresponding Voronoi tiling. Forces on the cells are computed using the connectivity information of the Voronoi tiling and activity is modelled as a self-propelling force in the direction set by a polarity vector assigned to each cell \cite{bi2016motility,barton2017active,huang2023bridging}. These models can efficiently capture cell rearrangement \cite{barton2017active,teomy2018confluent}, but are restricted by the assumption that the tissue can be represented as a Voronoi tiling, which in many situations is not the case.

Two-dimensional vertex models represent the apical side of an epithelial tissue as a polygonal tiling of a plane. Each cell is represented as a polygon, with two cells sharing an edge (i.e. a junction). Three or more junctions meet at a vertex, which acts as the degree of freedom \cite{honda1980much,farhadifar2007influence,fletcher2014vertex,sego2023general}. This allows modelling changes of cell shape due to mechanical forces \cite{bi2015density,alt2017vertex}.

Several studies have used vertex models to explore tissue homeostasis and its mechanical underpinnings. For example, Refs.\ \cite{farhadifar2007influence,kursawe2015capabilities} examined how cell mechanics, cell-cell interactions, and proliferation influence epithelial packing, demonstrating the critical roles these factors play in determining cell-packing geometries and transitions during tissue growth. Additionally, the mechanical behaviour of epithelial monolayers under external stretch highlighted how cell division contributes to tissue plasticity and stress adaptation, thereby providing insights into tissue viscoelasticity \cite{xu2015changes}. Furthermore, the morphodynamics of growing epithelial tissues have been shown to be significantly impacted by mechanical stresses, which play a crucial role in regulating tissue growth \cite{lin2017dynamic}. In contrast to these studies, here we examine the combined effect of the mechanical properties of the cells and the confinement on epithelial tissue homeostasis.

This paper is organised as follows: In Sec.~\ref{sec:Model}, we provide details of the two-dimensional vertex model for epithelial tissue mechanics, including the implementation of cellular processes within this framework. In Sec.~\ref{sec:results}, we present our findings by characterising homeostasis in terms of the variation in the (i)~number of active cells, (ii)~area occupied by active cells, (iii)~polygonal distribution of the active cell shapes, (iv)~realised shape index (i.e. the ratio of cell perimeter to the square root of its area) of the active tissue, and (v)~the cell pressure.Specifically, Sec.~\ref{sec:temporal_dynamics} describes the temporal dynamics as the system progresses towards homeostasis. Then, in Sec.~\ref{sec:modelp_effect}, we discuss the impact of model parameters, including growth rate, cell division, and cell ingression probability coefficients, on homeostasis, and in Sec.~\ref{sec:p0s_effect} we discuss the sensitivity of the homeostatic state to the target shape indexes of active and passive cells. In Sec.~\ref{sec:conf_effect}, we present the analysis of the effect of confinement on homeostasis. Lastly, summary and conclusions are given in Sec.~\ref{sec:conc}.

\section{Model}
\label{sec:Model}
\subsection{Two-dimensional vertex model for epithelial tissue mechanics}

Epithelial cells are tightly packed to form a confluent monolayer (e.g. epiblast of amniote embryos, the lining of blood vessels, kidney tubules, intestine, etc.) or a multilayer (e.g. skin, excretory ducts of sweat glands, etc.) sheet~\citep{betts2023epithelial}. The tissue cohesion is achieved by an adhesion belt formed of clusters of E-cadherin molecules concentrated in the adherens junctions~\citep{guillot2013mechanics}, giving epithelial tissues viscoelastic properties. While the shapes of cells in an epithelial monolayer generally resemble prisms, the molecules primarily responsible for the generation and transmission of mechanical forces are located close to the apical side (i.e. the top surface) of the cells. Therefore, a common approach is to focus on the apical side alone and approximate the tissue as a two-dimensional polygonal tiling. Cells are thus represented as polygons that share junctions, and three or more junctions meet at a vertex. This description is known as the two-dimensional vertex model~\citep{honda1978description, farhadifar2007influence, fletcher2014vertex}.

The mechanical energy of the model epithelium is determined by cell shapes, and it is given as
\begin{equation}
    E =  \sum_{c \, \in \, \text{cells}} \left[\frac{K_{A}}{2}(A_c-A_{c,0})^2+\frac{K_{P}}{2}(P_c-P_{c,0})^2\right], \label{eq:vm-bare}
\end{equation}
where the sum is over all the cells in the tissue. The first term is the penalty associated with changes in the cell's area, $A_c$, from a reference value $A_{c,0}$, and it accounts for the volume conservation and preferred height of actual cells. The second term penalises deviations in the cell's perimeter, $P_{c}$, from a reference value, $P_{c,0}$, and models the mechanical properties of the adhesion belt. The associated elastic moduli are $K_A$ and $K_P$, respectively. The ratio $p_{c,0}=P_{c,0}/\sqrt{A_{c,0}}$ is called the \emph{target cell shape index}. Finally, we have omitted the cell dependence of $K_A$ and $K_P$ to declutter the notation. Throughout this work, we use ``simulation'' units chosen such that the typical simulation box has size $L=50$ and the perimeter modulus $K_P=1$.

If all cells have the same target areas, $A_0$, and perimeters, $P_0$, the target shape index $p_{0} = P_{0}/\sqrt{A_{0}}$ determines the tissue's mechanical behaviour, distinguishing fluid--like and solid--like responses. Once $p_0$ exceeds a critical value, the shear modulus of the tissue decreases to zero, and the tissue fluidises. This is accompanied by the disappearance of the energy barrier for cell neighbour exchanges~\citep{bi2014energy}. For a tissue made entirely of hexagonal cells, the shear modulus falls to zero at the critical value $p_0 = 6/\sqrt{3\sqrt{3}/2}\approx3.722$, i.e. the perimeter to the square root of the area ratio of a regular hexagon. However, the energy barrier for neighbour exchanges remains finite until $p_0 \approx 3.81$, the corresponding ratio of a regular pentagon~\citep{sahu2020linear}. For random tilings, the solid-fluid transition has been reported to be in the range $p_0\approx 3.81-3.94$~\citep{bi2015density,merkel2019minimal,wang2020anisotropy,tong2022linear}. Values of $p_0\gtrsim{4.1}$ often lead to unstable simulations due to irregular cell shapes and overlap. Therefore, the relevant range of values of $p_0$ is generally between $\approx3.6$ and $\approx 4.0$. In this study, we varied $p_0$ within the range of $3.600-3.825$ because simulations become unstable for $p_0 > 3.825$.

\begin{figure}[t]
  \centering
  \includegraphics[width=0.6\textwidth]{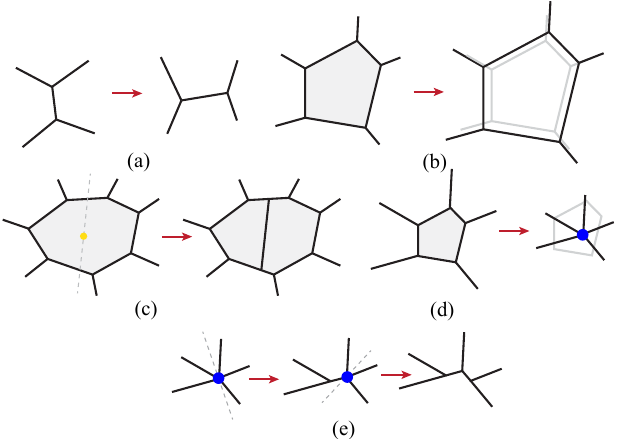}
  \caption{\label{fig:model} Five morphological changes of the model tissue. (a)~Intercalation (i.e. cell neighbour exchange) is implemented via a T1 transition where a junction shorter than a threshold length $\ell_{T1}$ is rotated $90^\circ$ counterclockwise and vertex connectivity is updated. (b)~Cell growth is implemented by rescaling the target area $A_{c,0}$ [Eqn.~\eqref{eq:cell-growth}]. (c)~Cells are divided perpendicular to their long axis along a line that passes through the centroid with a probability that is a function of the cell size [Eqn.~(\ref{eq:cell-div})]. (d)~Cells are removed with a probability that is a function of the cell size [Eqn.\ (\ref{eq:ingress})], by collapsing cell edges into a single vertex.  (e)~Vertices with four or more neighbours are resolved by spitting them at random while ensuring that mesh connectivity is not violated.}
\end{figure}

Cells in epithelial tissues are dynamic, but their motion is slow (typically, $\sim 10\,$\textmu m/h). Therefore, the motion is overdamped and can be approximated as a force balance between dissipative and mechanical forces. Under the assumption that cells are supported by a substrate that provides most of the dissipation via  friction, the equation of motion for the vertex $i$ at the position $\mathbf{r}_i$ is
\begin{equation}
    \zeta \dot{\mathbf{r}}_i = \mathbf{F}_i. \label{eqn:EOM}
\end{equation}
The overdot represents the time derivative,  $\zeta$ is the friction coefficient, and $\mathbf{F}_i$ is the force on the vertex. In simulation units, $\zeta=1$. If only passive mechanical forces are present, $\mathbf{F}_i=-\nabla_{\mathbf{r}_i}E$, where $\nabla_{\mathbf{r}_{i}}$ is the gradient with respect to the position vector $\mathbf{r}_i$ of vertex $i$, and mechanical energy $E$ is defined in Eqn.~\eqref{eq:vm-bare}. In the presence of activity, $\mathbf{F}_i$ can be extended to include additional force contributions~\citep{henkes2020dense,sknepnek2023generating}. Finally, the noise term that would normally accompany the dissipative term in the equation of motion~\cite{zwanzig2001nonequilibrium} is omitted as is usual in vertex models, since the effects of the microscopic noise on the long-time behaviour of cells in epithelial tissue are usually negligible. 

To maintain proper function, epithelial cells rearrange via neighbour exchanges, grow, divide, and die. Therefore, the vertex dynamics needs to be augmented to include cellular processes such as intercalations (i.e. cell neighbour exchanges), ingression/extrusion (i.e. removal of individual cells from the tissue), cell division, and cell growth (Fig.~\ref{fig:model}). Modelling these processes requires updates of the connectivity of the polygonal tiling via topological changes such as T1 (intercalation) and T2 (ingression/extrusion) transitions~\citep{fletcher2013implementing}.

\subsection{An active inclusion in a passive tissue patch}
\label{sec:2.2}

The features of the vertex model discussed above are general. We now proceed to specific extensions of the model used to study the proliferation of an active patch embedded in a passive tissue. Since we are interested in a general understanding of the role of mechanical interactions between the proliferating tissue and its environment in maintaining homeostasis, we chose a simple square geometry. Therefore, we study a square patch of size $L_\text{a}\times L_\text{a}$ of a proliferating active epithelial tissue embedded in a passive epithelium confined to a square box of size $L\times L$. The passive tissue is clamped to the edges of the box, as illustrated in Fig.~\ref{fig:time_evol}a. Passive cells can move and rearrange, but cannot grow, divide, or be extruded. Active cells, however, can grow, divide, and be extruded. 

We initialise the simulation with a disordered tiling generated by placing $N$ points at random in a square box. The points act as the initial seeds for a Voronoi tessellation. Upon building the Voronoi tesselation, the seed points are moved to the centroids of Voronoi tiles, and a new Voronoi tiling is constructed. The procedure is repeated iteratively until the maximal relative difference between positions of seed points in two consecutive iterations is below $5\cdot10^{-5}$. This results in a centroidal Voronoi tesselation, which has the property that the tiling is random, but all cells are of similar size and shape~\citep{du1999centroidal}. Additionally, we sample the target perimeters of cells in the active tissue ($P_{c,0}^\text{a}$) from a normal distribution with a mean of $\mu=5.98$ and a standard deviation of $\sigma=0.3$. The corresponding target areas of the active cells are determined by $A_{c,0}^\text{a} = (P_{c,0}^\text{a}/p_0^\text{a})^2$, where $p_0^{\text{a}}$ is the target shape index of active cells. For the passive cells, the target perimeters and areas are set to constant values $P_{0}^\text{p} = 5.98$ and $A_{0}^\text{p} =(P_{0}^\text{p}/p_0^\text{p})^2$, respectively. Furthermore, the vertices of the outermost layers are fixed in position and constrained to lie along a straight line, reflecting a clamped boundary condition \citep{ustinov2022elastic}. Straight boundaries are created during the initial configuration setup by padding the simulation box and mirroring seed points within a preset cutoff distance from the boundaries.

Before starting the full simulation, we perform passive relaxation, i.e.\ each random initial configuration is evolved for time $t_\text{rel}=50$ without cell growth, divisions, and ingressions. This allows the model tissue to relax close to a local energy minimum and prevents large vertex movements in the initial stages of the actual simulation. Unless stated otherwise, simulations were performed from a single initial configuration.

Cell intercalations are implemented via a T1 transition (Fig.~\ref{fig:model}a) based on a minimum junction length~\cite{nagai2001dynamic}. If the junction length drops below a specified threshold, $\ell_\text{T1}$, it is rotated by $90^\circ$ counterclockwise and extended to the length $\ell_\text{T1}^\text{n}=1.02\ell_\text{T1}$. Then the connectivity is updated to account for the exchange of cell neighbours. Moreover, T1 transitions rarely result in either a passive cell entering the proliferating region or an active cell entering the passive region (i.e.\ a cell of one type is fully surrounded by cells of the other type). In these cases, the cell type is changed from passive to active, or vice-versa.

Next, we discuss how we implement the active processes, i.e.\ cell growth, division, and ingressions. For simplicity, we assume that the cell growth (Fig.~\ref{fig:model}b) is described by a linear model, where the target area of active cells, $A^\text{a}_{c,0}(t)$, evolves as
\begin{equation}
\dot{A}^\text{a}_{c,0} = g, \label{eq:cell-growth}
\end{equation}
where $g$ is a constant growth rate. The target shape index of active cells is kept constant at $p_0^\text{a}$ by adjusting the target perimeter to $P^\text{a}_{c,0}(t) = p_0^\text{a}\sqrt{A^\text{a}_{c,0}(t)}$. The target area and perimeter of all passive cells are assumed to be identical and time-independent, giving the target shape index $p_0^\text{p}$. 
  
Cells in the active region can divide, and the division mechanism is stochastic. The division probability, $\mathbb{P}_\text{d}(c)$, of a cell $c$ increases with the cell area $A_c$ as
\begin{equation}
    \mathbb{P}_\text{d}(c) = \frac{1}{1+e^{-\alpha(A_c-A_\text{d})}},\label{eq:cell-div}
\end{equation}
where $A_\text{d}$ represents the area at which the probability of cell division is $1/2$ and $\alpha$ regulates the sensitivity of division probability to changes in the cell area. $A_\text{d}$ is set to $1.6\bar{A}^\text{a}(t=t_\text{ref})$, where $\bar{A}^\text{a}(t=t_\text{ref})$ is the mean area of active cells after the initial passive relaxation for $t_\text{ref}$ steps. The division probability is calculated for each cell once every five simulation time steps, and cell division occurs if a uniformly distributed random number drawn from the interval $[0,1]$ is smaller than the calculated probability $\mathbb{P}_\text{d}(c)$. The division process follows Hertwig's rule~\citep{hertwig1884problem}, i.e. a cell is divided into two daughter cells by choosing a direction perpendicular to its long axis that passes through its centroid (Fig.~\ref{fig:model}c). The long axis is determined by diagonalising the gyration tensor~\citep{rozman2023shape}. The target perimeters of two daughter cells are separately drawn from the normal distribution with a mean of $\mu=5.98$ and standard deviation $\sigma=0.3$ and target areas are calculated as $A_{\text{daughter},0}^\text{a}=(P_\text{daughter,0}^\text{a}/p_0^\text{a})^2$. 

Similarly, for cell ingression, we define the probability $\mathbb{P}_\text{i}(c)$ which increases with a decrease in cell area $A_c$ as,

\begin{equation}
    \mathbb{P}_\text{i}(c) = \frac{1}{1+e^{\beta(A_c-A_\text{i})}},\label{eq:ingress}
\end{equation}
where $A_\text{i}$ represents the area at which the probability of cell ingression is $1/2$ and $\beta$ regulates the sensitivity of ingression probability to changes in the cell area. $A_\text{i}$ is set to $0.3\bar{A}^\text{a}(t=t_\text{ref})$. Ingression of a cell typically leads to the formation of a vertex with more than three neighbours (Fig.~\ref{fig:model}d). Such vertices are resolved by picking a random ``cut'' direction that does not violate the tissue connectivity and inserting a new edge of length $\ell_\text{T1}^\text{n}$. The procedure is repeated until all high-coordination vertices are resolved (Fig.~\ref{fig:model}e).  

These five morphological transformations of the model tissue are shown schematically in Fig.\ \ref{fig:model}. Finally, the equations of motion [Eqn.~(\ref{eqn:EOM})] are solved using the first-order Euler method \cite{barton2017active,yamamoto2022non,lin2023structure,rozman2023shape,rozman2023dry} with a time step $\delta t$, implemented in an in-house developed software package called the Active Junction Model (AJM) \cite{AJM_git}, specifically designed for simulating the vertex model. Unless otherwise specified, the parameters and the values used are listed in Table~\ref{tab:vertex_params}. Notably, the target shape indices of active and passive cells are varied in the range of $p_0^\text{p},p_0^\text{a}\in[3.600,3.825]$. The parameters $l_{\text{T1}}$, $\delta t$, and $K_A$ are simulation parameters chosen to optimize the balance between numerical stability and simulation time. The computational time for a single simulation depends on the system size, but typical runs take between 48 to 72 hours on a single core of a 32-core 2.35GHz AMD EPYC 7452 CPU.

\begin{table}[tbh]
\caption{Values of the parameters in simulation units.}
\label{tab:vertex_params}
\centering
\begin{tabular}{lll}
\hline\noalign{\smallskip}
Parameter & Description & Numerical Value  \\
\noalign{\smallskip}\hline\noalign{\smallskip}
${K}_{A}$ & Area elastic modulus &   3.0 \\
$N$ & Initial number of cells & $10^3$ \\
$P_0^\text{p}$ & Target perimeter of passive cells & 5.98 \\
$\bar{P}_0^\text{a}$ & Mean initial target perimeter of active cells & 5.98 \\
$\sigma$ & Standard deviation of active cells target perimeter  &  $0.3$ \\
$g$ & Growth rate &  $2.0\cdot10^{-3}$ \\
$A_\text{d}$ &  Area at which $\mathbb{P}_\text{d}(c) = 0.5$ &$1.6 \ \bar{A}^{\text{a}}(t=50)$ \\
$A_\text{i}$ & Area at which $\mathbb{P}_\text{i}(c) = 0.5$ &  $0.3 \ \bar{A}^{\text{a}}(t=50)$ \\
$\ell_\text{T1}$ & T1 transition threshold & $5\cdot10^{-3}$ \\
$\delta t$ & Simulation time step & $5\cdot10^{-3}$ \\
$\alpha$ & Division probability parameter [Eqn.~\eqref{eq:cell-div}] & $8.0$ \\
$\beta$ & Ingression probability parameter [Eqn.~\eqref{eq:ingress}] &  $6.0$ \\
$p_0^\text{a}$ & Target shape index of active cells & $3.600-3.825$ \\ [0.25ex]
$p_0^\text{p}$ & Target shape index of passive cells &  $3.600-3.825$ \\
$L_\text{a}$ & Initial size of the active patch &  $20.0$ \\
$t_\text{rel}$ & Initial configuration relaxation time & $50$ \\

\hline\noalign{\smallskip}
\end{tabular}
\end{table}

\section{Results and Discussion}
\label{sec:results}

\subsection{Temporal dynamics of the tissue}
\label{sec:temporal_dynamics}
\begin{figure}[tbh]
  \centering
  \includegraphics[width=0.9\linewidth]{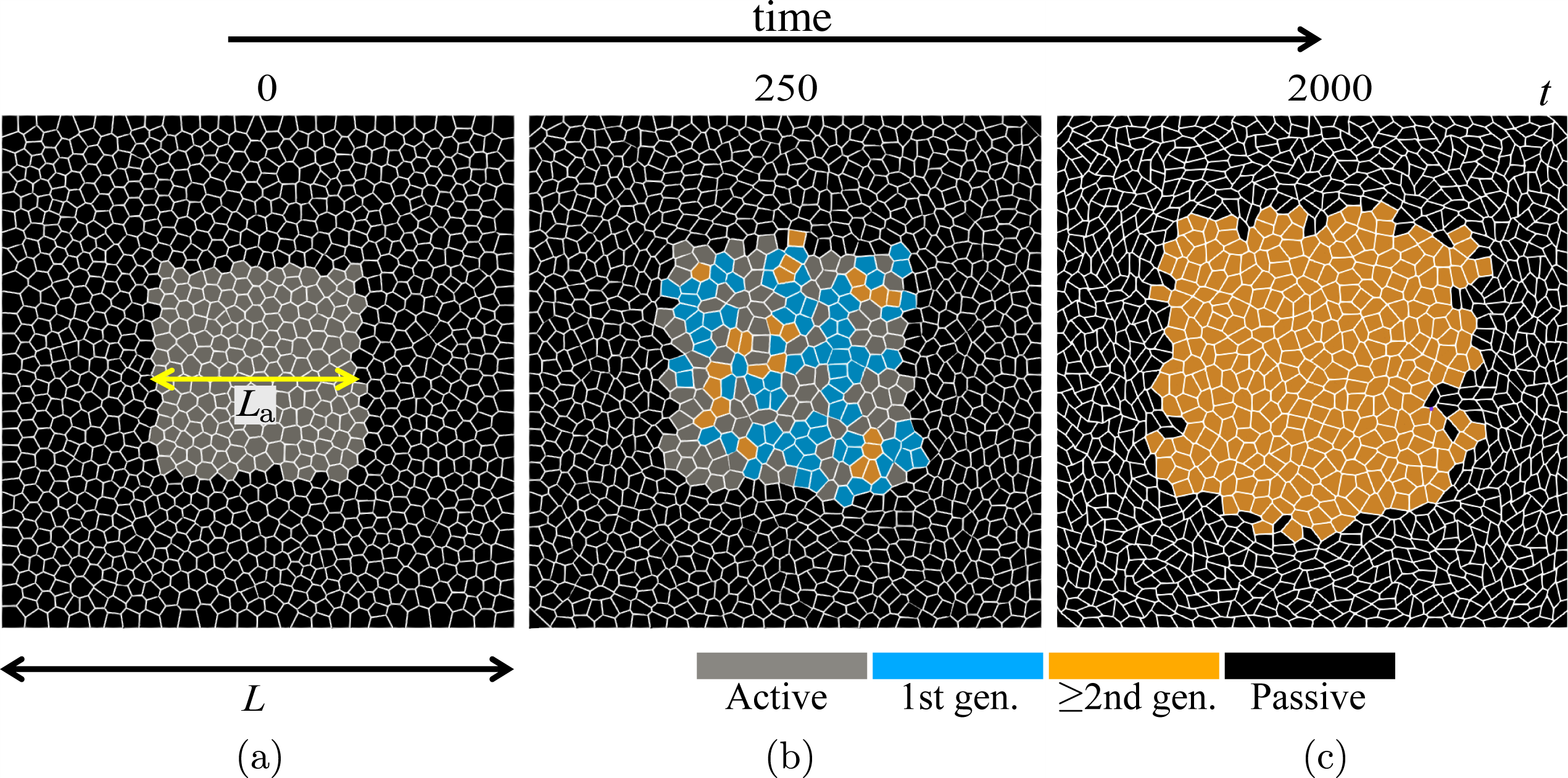}
  \caption{\label{fig:time_evol} 
The snapshots of the evolution of the model tissue: (a) initial configuration with active inclusion containing $169$ cells (light grey) surrounded by passive tissue with $831$ cells (black); (b) intermediate configuration showing cells born in the active region after the first division ($1^\text{st}$ generation - blue) and subsequent divisions ($\geq2^\text{nd}$ generation - orange); and (c) a steady-state configuration with all initial active cells replaced by new generation cells. These snapshots correspond to $p_0^\text{a} = 3.60$ and $p_0^\text{p} = 3.80$.}
\end{figure}

We first study the time evolution of the tissue towards homeostasis, as depicted in Fig.~\ref{fig:time_evol}. Cells in the central proliferating region grow, divide, ingress, and rearrange. Consequently, the region expands and exerts compressive stress on the passive cells, leading to their deformation and intermittent rearrangements. Constrained by the fixed boundary of the simulation box, the passive cells get compressed. Eventually, the system reaches a steady state (i.e. homeostasis), characterised by continuously replenishing cells in the proliferating region. 

\begin{figure}[tbh]
  \centering
  \includegraphics[width=\linewidth]{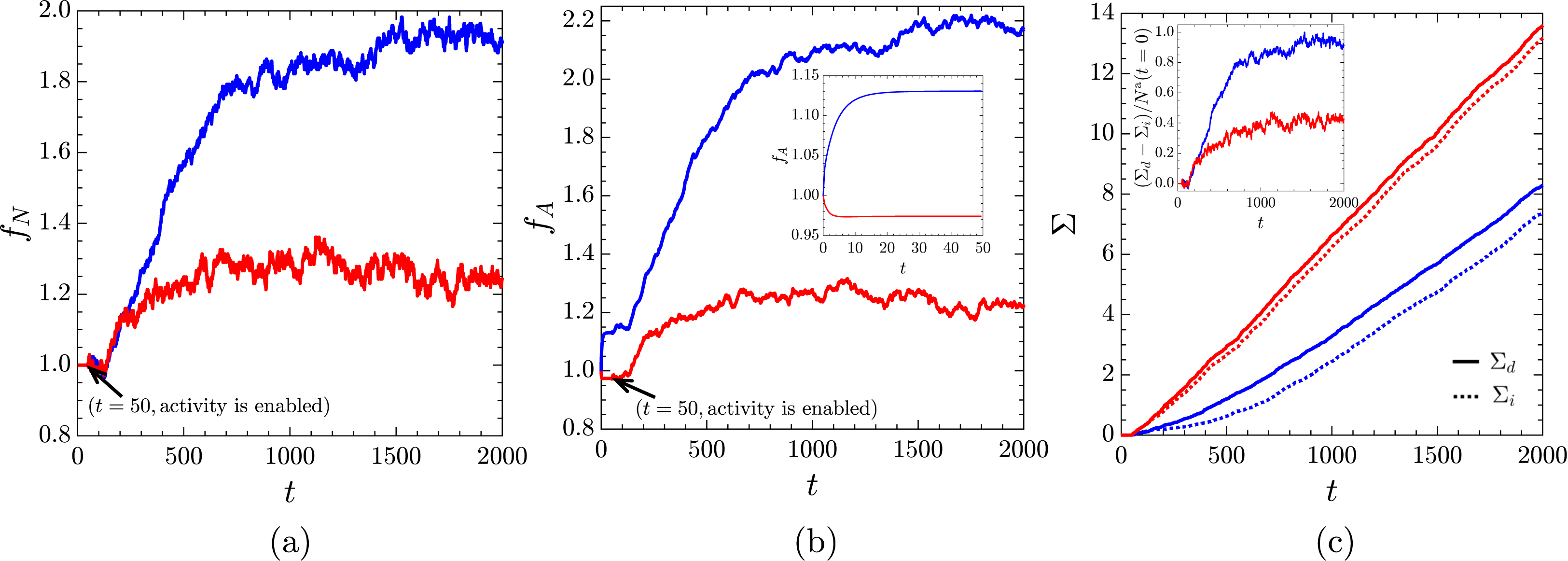}
  \caption{\label{fig:time_evol_Nf_sigma} (a) The time dependence of the fractional change in the number of active cells ($f_N$) defined in Eqn.~(\ref{eq:Fn}). (b) The fractional change in the area of active cells ($f_A$) defined in Eqn.~(\ref{eq:Fn}); Inset: Passive relaxation (i.e.\ growth, division, and ingression switched off). (c) The cumulative sum (\textSigma) of cell divisions (\textSigma$_d$ - solid lines) and ingressions (\textSigma$_i$ - dotted lines) normalized by $N^\text{a}(t=0)$; Inset: Time dependence of the difference between the cumulative sums of cell divisions and ingressions, \textSigma$_d$ $-$ \textSigma$_i$, normalized by $N^\text{a}(t=0)$. $p_0^\text{a} = 3.60$, $p_0^\text{p} = 3.80$ (blue curves) and $p_0^\text{a} = 3.80$, $p_0^\text{p} = 3.60$ (red curves). In panel (b), the data is normalised with respect to the reference area at $t_\text{ref}=0$, i.e. before the passive relaxation.}
\end{figure}

Figure~\ref{fig:time_evol_Nf_sigma} shows the time evolution of the fractional change in the number and area of active cells,
\begin{equation}
    f_N(t) = \frac{N^\text{a}(t)}{N^\text{a}(t=t_\text{ref})} \quad \text{and} \quad f_A(t) = \frac{A^\text{a}(t)}{A^\text{a}(t=t_\text{ref})}, \label{eq:Fn}
\end{equation}
where $N^\text{a}(t)$ and $A^\text{a}(t)$ are the number and area of active cells at time $t$, respectively, and $t_\text{ref}$ is the reference time used for normalisation, depending on the specific case, set either to $t_{\text{ref}} = 0$ (before passive relaxation) or to $t_{\text{ref}} = 50$ (after passive relaxation). While this distinction helps separate the contribution to $f_A$ from active processes and passive relaxation, $f_N$ remains insensitive to the choice of $t_\text{ref}$ because the number of cells does not change during relaxation. The cumulative sums of cell divisions (\textSigma$_d(t)$) and ingressions (\textSigma$_i(t)$) are also calculated as the system reaches homeostasis. The active cell population gradually increases and reaches a dynamic steady state where both $f_N$ and $f_A$ saturate but continue to fluctuate around a mean value (Fig.~\ref{fig:time_evol_Nf_sigma}a,b). Notably, the changes in the area due to active processes (i.e.\ growth, divisions, and ingressions) are significantly larger than those due to passive relaxation to mechanical equilibrium compatible with a given shape index and the target area (Fig.~\ref{fig:time_evol_Nf_sigma}b - inset).

The homeostatic state is characterised by a dynamical balance between divisions and ingressions (Fig.~\ref{fig:time_evol_Nf_sigma}c), i.e. the cumulative sums of cell divisions and ingressions continuously grow but retain, on average, a constant difference (Fig.~\ref{fig:time_evol_Nf_sigma}c - inset). The homeostasis is, therefore, achieved through a balance of cell proliferation and cell removal~\cite{kaliman2021mechanical,o2022tissue}. This dynamic balance is crucial for maintaining tissue integrity and function, preventing the tissue from overgrowing or collapsing, as highlighted in experimental studies  \cite{eisenhoffer2012crowding,kaliman2021mechanical}.

Furthermore, the steady-state values of $f_N$ and $f_A$ depend on the mechanical properties of the tissue, specifically the target shape indices of active ($p_0^\text{a}$) and passive ($p_0^\text{p}$) cells. Since $p_0\propto(A_0)^{-1/2}$, and we control $p_0$ by tuning $A_0$, increasing $p_0$ reduces preferred cell size. As the cell area contribution to energy is $\propto(A-A_0)^2$, cells with smaller $A_0$ are easier to compress to the same $A$, i.e.\ are effectively softer.  When $p_0^\text{a} = 3.60$ and $p_0^\text{p} = 3.80$ (i.e.\ stiffer and larger active cells surrounded by softer and smaller passive cells), the increase in active cell count and the area they occupy is $\approx1.5$ times higher compared to the scenario where $p_0^\text{a} = 3.80$ and $p_0^\text{p} = 3.60$ (i.e.\ softer and smaller active cells surrounded by stiffer and larger passive cells). This suggests that a stiffer proliferating tissue composed of larger cells enclosed by a softer passive tissue of smaller cells reaches a homeostatic state with more active cells, which occupy a significantly larger total area than their initial area.

Figure~\ref{fig:time_evol_Nf_sigma}c shows that the target shape indices of active ($p_0^\text{a}$) and passive ($ p_0^\text{p}$) cells play an important role in determining cell proliferation and net population changes. A configuration with a stiffer proliferating tissue surrounded by a softer passive tissue ($p_0^\text{a} = 3.60$ and $p_0^\text{p} = 3.80$) results in fewer cell divisions, but the difference between the number of divisions and the number of ingressions is larger, leading to a greater net gain of active cells. Conversely, reversing this configuration ($p_0^\text{a} = 3.80$ and $p_0^\text{p} = 3.60$) leads to more cell divisions but also a higher number of ingressions, resulting in a smaller net gain of active cells. This suggests that the tissue can adapt to its surroundings to maintain stability and proper function~\cite{vogel2006local,humphrey2014mechanotransduction}.

These findings highlight the delicate balance between cell divisions and ingressions in maintaining tissue homeostasis. A configuration with higher division rates does not necessarily lead to a larger active cell population if it is accompanied by a proportionate increase in ingressions. Instead, the optimal balance, as seen with \( p_0^\text{a} = 3.60 \) and \( p_0^\text{p} = 3.80 \), results in a more effective net gain of active cells, underscoring the importance of mechanical properties in regulating tissue dynamics.

 \subsection{Effects of the growth rate, and division and ingression probability coefficients}
 \label{sec:modelp_effect}

 \begin{figure}[tbh!]
\centering
  \includegraphics[width=\textwidth]{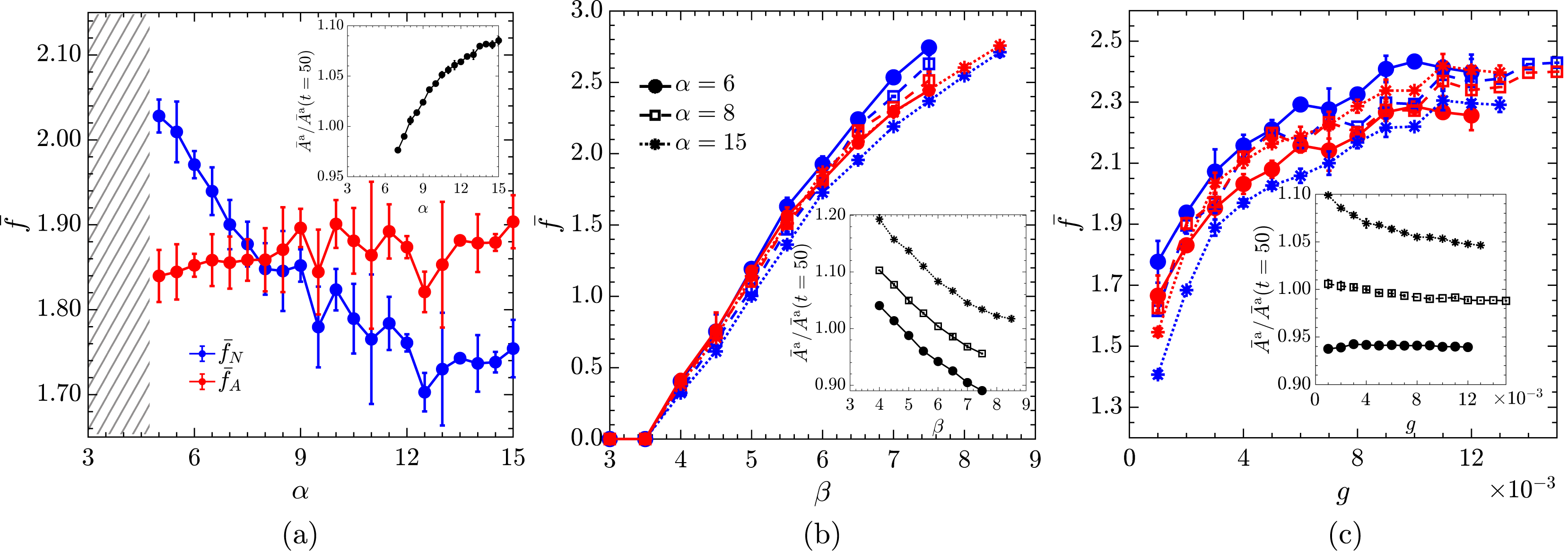}
\caption{\label{fig:effect_of_params} The mean fractional change in the number of active cells ($\bar{f}_N$; blue) and the area occupied by the active cells ($\bar{f}_A$; red) as a function of (a) division probability coefficient, $\alpha$ for $\beta = 6$ and $g = 0.002$, (b) ingression probability coefficient, $\beta$ for $\alpha = 6$ (solid), $\alpha=8$ (dashed), and $\alpha=15$ (dotted) and $g = 0.002$, and (c) growth rate, $g$, for $\alpha = 6$ (solid), $\alpha=8$ (dashed), and $\alpha=15$ (dotted) and $\beta = 6$. The error bars represent one standard deviation of the mean values from three different realisations. The shaded region in (a) indicates parameter values that lead to unphysical configurations with overlapping cells. In (b) and (c), the abrupt end of the curves, similarly, indicates that beyond those parameter values, simulations become unstable. The insets in (a), (b), and (c) show the variation in the mean area of active cells normalized to their mean area at $t = 50$ with $\alpha$, $\beta$, and $g$, respectively. $p_0^\text{a} = 3.60$ and $p_0^\text{p} = 3.80$ in all plots.}
\end{figure}

Next, we investigate the effects of the growth rate ($g$) and the parameters that control the shape of the probabilities of division ($\alpha$) and ingression ($\beta$) on the homeostatic state for $p_0^\text{a} = 3.60$ and $p_0^\text{p} = 3.80$. This configuration was selected since the difference in the number of cell divisions and ingressions is high, leading to the maximum number of active cells, as illustrated in Fig.~\ref{fig:time_evol_Nf_sigma}.

We characterise the homeostatic state in terms of the steady-state mean values of the fractional change in the number ($\bar{f}_N$) and the area ($\bar{f}_A$) of active cells defined, respectively, as
\begin{equation}\label{eq:mean_change}
   \bar{f}_N=\langle f_N(t_i)\rangle_t \quad \text{and} \quad \bar{f}_A=\langle f_A(t_i)\rangle_t.
\end{equation}
Here, the time averaging $\langle\dots\rangle_t$ is done for the time interval $\Delta t=500$. The averaging is started after the system reaches homeostasis, which is typically between t=1500 and t=2000 (Appendix~\ref{appA}).

Figure~\ref{fig:effect_of_params} shows the dependence of the mean fractional changes of $\bar{f}_N$ and the area occupied by active cells on the division probability coefficient ($\alpha$), the ingression probability coefficient ($\beta$), and the growth rate ($g$).

While $\bar{f}_N$ decreases with increasing $\alpha$, $\bar{f}_A$ remains approximately constant (Fig.~\ref{fig:effect_of_params}a). The division probability curve is steeper for large values of $\alpha$, reducing the likelihood of smaller cells dividing. As a result, there is a significant decrease of $\bar{f}_N$ and $\bar{f}_N < \bar{f}_A$ for $\alpha>8$. At $\alpha \approx 8$, the $\bar{f}_N$ and $\bar{f}_A$ curves intersect, corresponding to the total increase in the area occupied by active cells closely matching the increase in the total number of those cells. Therefore, the mean area per active cell remains close to the mean reference area before the activity was turned on, i.e.\ $\bar{A}^\text{a}\approx\bar{A}^\text{a}(t_\text{ref})$. For $\alpha < 8$, divisions are more frequent and involve smaller cells, resulting in $\bar{f}_N > \bar{f}_A$, i.e.\ the mean cell area decreases. Since $\bar{f}_A$ does not change significantly, the mean area of active cells increases with $\alpha$ (inset of Fig.~\ref{fig:effect_of_params}a), indicating that active cells adjust their areas accordingly.

On the other hand, both $\bar{f}_N$ and $\bar{f}_A$ increase with $\beta$ (Fig.~\ref{fig:effect_of_params}b). Higher values of $\beta$ make the ingression probability curve steeper, decreasing the likelihood of cells ingressing at higher areas. This substantially increases $\bar{f}_N$ and $\bar{f}_A$, with their values weakly depending on $\alpha$. As $\beta$ increses, ingressions can no longer compensate cell divisions and the system becomes unstable.  Furthermore, for a fixed $\beta$, the mean area of active cells, $\bar{A}^\text{a}$, increases with an increase in $\alpha$ (Fig.~\ref{fig:effect_of_params}b - inset), consistent with Fig.~\ref{fig:effect_of_params}a.

Finally, Fig.~\ref{fig:effect_of_params}c shows that both $\bar{f}_N$ and $\bar{f}_A$ increase with the growth rate, $g$, until the tissue becomes unstable. For a fixed $g$, the values of $\bar{f}_N$ and $\bar{f}_A$ depend on $\alpha$ following the same trend as in Fig.~\ref{fig:effect_of_params}a. On the other hand, the mean cell area of active cells depends weakly on $g$, remaining nearly constant for $\alpha=6$, and slightly decreasing with $g$ for higher alpha (Fig.~\ref{fig:effect_of_params}c - inset). Additionally, for a given $g$, the mean area of active cells increases with $\alpha$, consistent with Fig.~\ref{fig:effect_of_params}a.

 \subsection{Effects of target shape index of active and passive cells on the homeostatic state}
 \label{sec:p0s_effect}

\begin{figure}[tbh]
  \centering
  \includegraphics[width=0.66\linewidth]{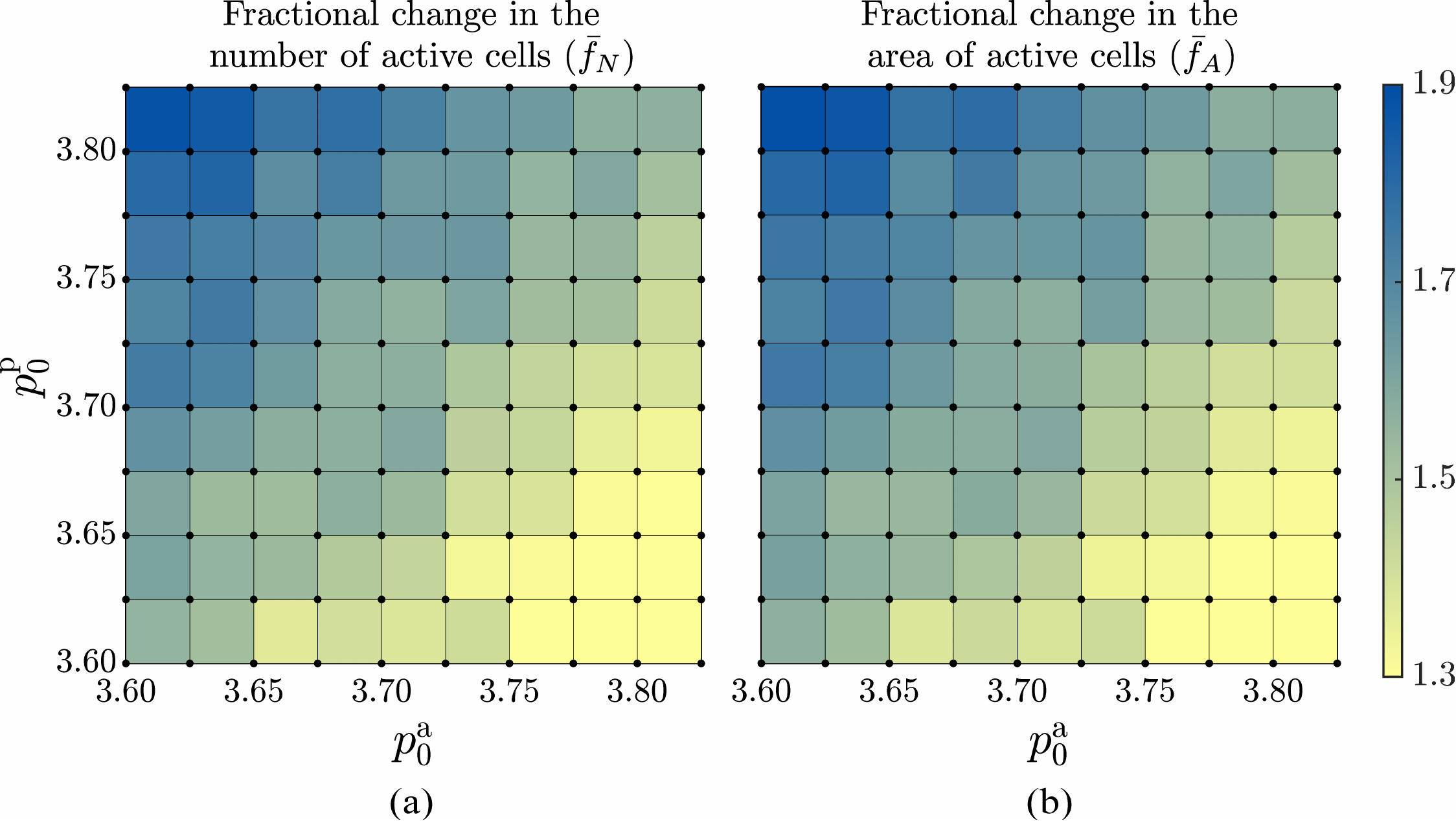}
\caption{\label{fig:nf_af_state_diag}
(a) The fractional change of the average number of active cells, $\bar{f}_N$, and (b) the average area occupied by the active cells, $\bar{f}_A$, [Eqns.~(\ref{eq:mean_change})] in the steady state as a function of the target shape index of active ($p_0^\text{a}$) and passive ($p_0^\text{p}$) cells. In panel (b) the data is normalised with respect to the configuration after passive relaxation ($t_\text{ref}=50$). The colour bar applies to both panels. Simulations are performed at grid points indicated by black circles, and the colour within each square represents the average of the values at the four corners of that square. The standard deviation of the time series used to compute the mean values is within the range of $0.01-0.05$. } 
\end{figure}

Next, we examine the effect of target shape indices of active and passive cells on the homeostatic state for $\alpha = 8$, $\beta = 6$, and $g = 0.002$. This configuration corresponds to the values of parameters where $\bar{f}_N$ and $\bar{f}_A$ curves cross (Fig.~\ref{fig:effect_of_params}), i.e.\ when active cells nearly retain their original area making simulations stable over the full parameter range. Figure \ref{fig:nf_af_state_diag}a shows that the dependence of the final cell count of active cells measured in terms of $\bar{f}_N$ on the target shape indexes of both passive, $p_0^\text{p}$, and active, $p_0^\text{a}$, cells. The mean fractional change of the number of active cells, $\bar{f}_N$, increases as $p_0^\text{p}$ is increased from $3.600$ to $3.825$, for a given value $p_0^\text{a}$. Conversely, $\bar{f}_N$ decreases with increasing $p_0^\text{a}$. The dependence of the mean steady area fraction of active cells, $\bar{f}_A$, on $p_0^\text{a}$ and $p_0^\text{p}$, is shown in Fig.~\ref{fig:nf_af_state_diag}b. Not surprisingly, it closely matches the trend of $\bar{f}_N$ in Fig.~\ref{fig:nf_af_state_diag}a so that, irrespective of $p_0^\text{a}$ and $p_0^\text{p}$, $\bar{f}_N \approx \bar{f}_A$, consistent with the observation that $\alpha$, $\beta$ and $g$ primarily controls the relative variation of $\bar{f}_N$ and $\bar{f}_A$ (see Fig.\ \ref{fig:effect_of_params} and Appendix~\ref{appB}).

The increase of $\bar{f}_N$ and $\bar{f}_A$ as one moves from the bottom right to the top left corner in Fig.\ \ref{fig:nf_af_state_diag} can be understood for as follows. As the target shape index of the passive cells, $p_0^\text{p}$, increases, the target area of the passive cells decreases, and the passive tissue shrinks its size making more area available for active cells. As a result, the difference in the number of divisions and the number of ingressions increases (Appendix~\ref{appC}), leading to the increase of $\bar{f}_N$ and $\bar{f}_A$. Moreover, the dependence of the fractional change of both the number of active cells and the area they occupy on the target shape indices of active and passive cells remains qualitatively consistent, regardless of the specific values of model parameters $\alpha$, $\beta$, and the growth rate (Appendix~\ref{appB}).

\subsubsection{Homeostatic pressure in active and passive cells}

\begin{figure}[tbh!]
\centering
  \includegraphics[width=0.66\linewidth]{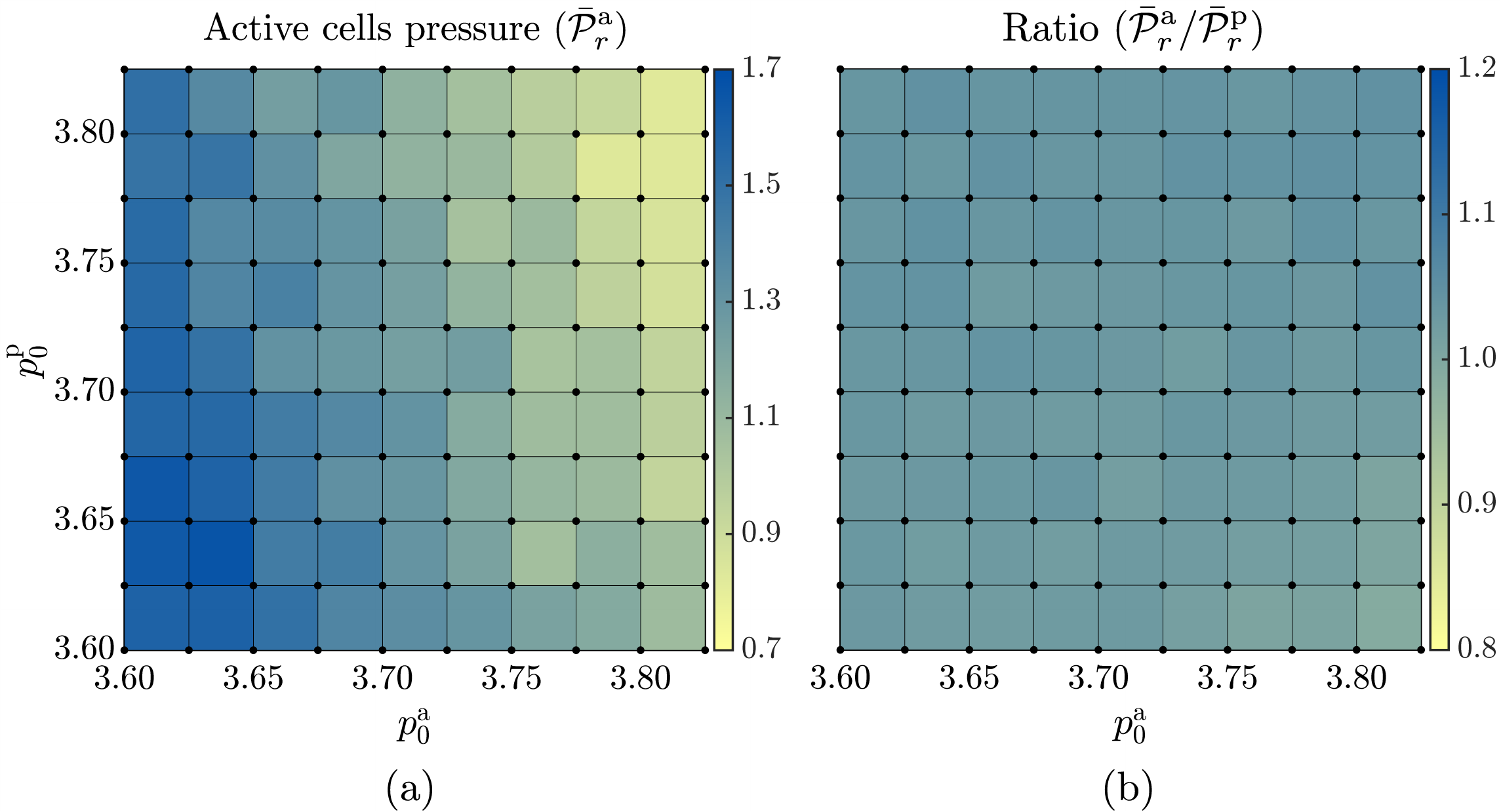}
\caption{\label{fig:pressure} The dependence of the (a) mean homeostatic pressure of active cells ($\bar{\mathcal{P}}_r^\text{a}$) and (b)  the ratio of the mean homeostatic pressure of active cells to that of passive cells ($\bar{\mathcal{P}}_r^\text{a}/\bar{\mathcal{P}}_r^\text{p}$) on the target shape indices of active, $p_0^\text{a}$, and passive, $p_0^\text{p}$, cells. Simulations are performed for parameter values corresponding to the grid points indicated by black circles, and the colour within each square represents the average of the values at the four corners of each square. The mean is calculated over $2 \cdot 10^4$ simulation time steps (i.e. time interval $\Delta t=500$ with the data recorded every five simulation steps) in the steady state.}
\end{figure}

Next, we examine the pressure in the active and passive cells in homeostasis. The pressure of a cell is related to the actual ($A_c$) and target ($A_{c,0}$) areas of the cell via \cite{tong2023linear},
\begin{equation}
    \mathcal{P}_{r}^c = -K_A (A_c-A_{c,0}).
\end{equation}

The mean pressure of active ($\bar{\mathcal{P}}_r^\text{a}$) and passive ($\bar{\mathcal{P}}_r^\text{p}$) cells is defined as
\begin{equation} \label{eqn:pressure}
    \bar{\mathcal{P}}_r^\text{a} = \left\langle\frac{1}{N^\text{a}(t_i)} \sum_{c \in \{\text{a}\}} \mathcal{P}_{r}^\text{c}(t_i)\right\rangle_t, \quad \bar{\mathcal{P}}_r^\text{p} = \left\langle\frac{1}{N^\text{p}(t_i)} \sum_{c \in \{\text{p}\}} \mathcal{P}_{r}^\text{c}(t_i)\right\rangle_t,
\end{equation}
where $N^\text{a}(t_i)$ and $N^\text{p}(t_i)$ are, respectively, the numbers of active and passive cells at time $t_i$, and $\langle\dots\rangle_t$ denotes the time average.

Figure~\ref{fig:pressure} shows the dependence of $\bar{\mathcal{P}}_r^\text{a}$ and the ratio $\bar{\mathcal{P}}_r^\text{a}/\bar{\mathcal{P}}_r^\text{p}$ on the target shape index of active and passive cells. The pressure in active cells (Fig.~\ref{fig:pressure}a) increases with a decrease in the target shape index of either cell type. The pressure ratio in active cells to the passive cells remains close to $1$ for all values of $p_0$ (Fig.~\ref{fig:pressure}b), indicating that the mechanical forces are balanced.

Finally, with pressure being proportional to the cell area, one might expect that the plot of the pressure of active cells (Fig.~\ref{fig:pressure}a) would have the same pattern as the mean area of active cells (Fig.~\ref{fig:nf_af_state_diag}b). This is, however, not the case since the target cell areas ($A_{c,0}$) also depend on the target cell shape index. 

\subsubsection{Cell neighbour count and disorder in the active tissue}

We proceed to quantify the active tissue in terms of the distribution of the cell neighbour count and characterise its disorder via the realised shape index of active cells. 

\begin{figure}[tbh!]
\centering
  \includegraphics[width=\linewidth]{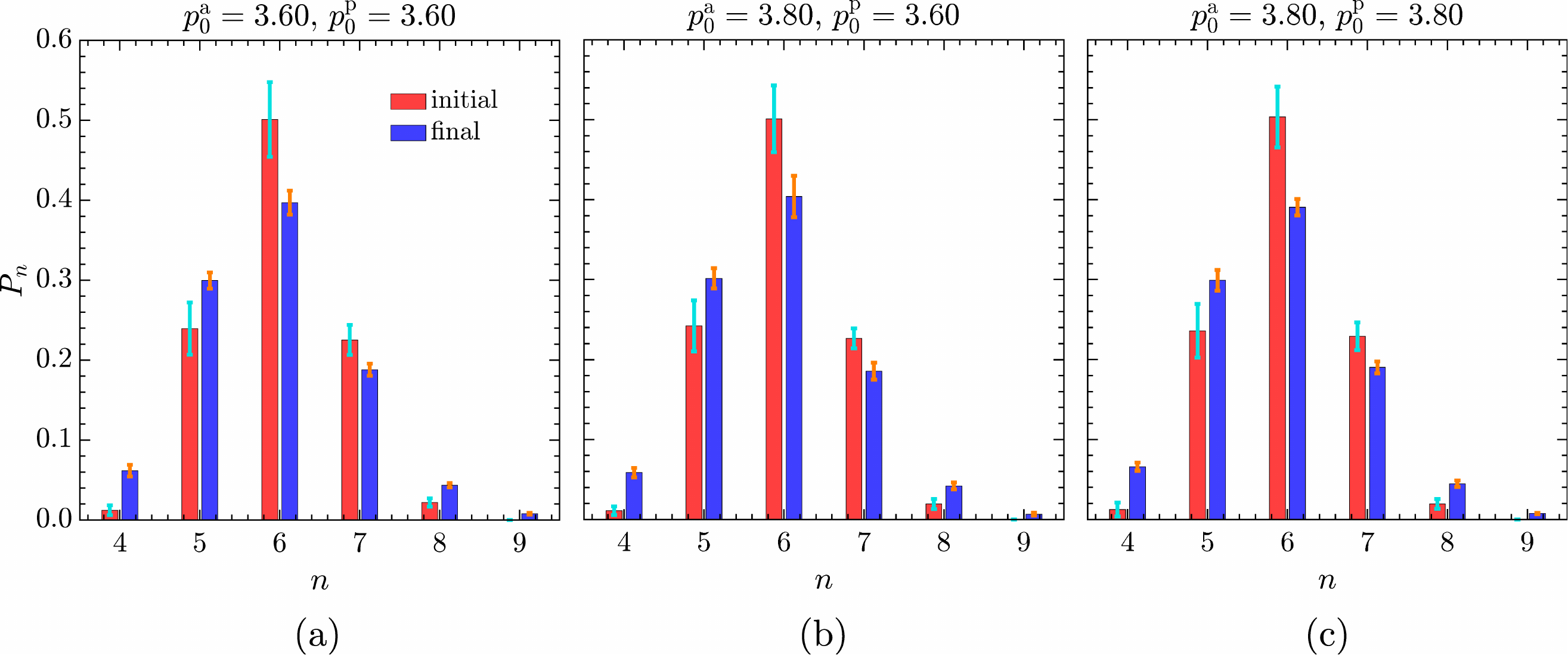}
\caption{\label{fig:polygonal_dist}Fraction of active cells with $n$ neighbours, $P_n$, averaged over time and across five different initial configurations for different target shape index combinations: (a) $p_0^\text{a} = 3.60$, $p_0^\text{p} = 3.60$, (b) $p_0^\text{a} = 3.80$, $p_0^\text{p} = 3.60$, and (c) $p_0^\text{a} = 3.80$, $p_0^\text{p} = 3.80$. The initial and final distributions, averaged across all realisations, are denoted by red and blue bars, respectively, with error bars indicating the corresponding standard deviations of the mean values. The mean value for each of the five initial configurations is calculated over the time interval $\Delta t = 500$.}
\end{figure}

Figure~\ref{fig:polygonal_dist} shows the mean distribution of cell neighbour counts averaged over time and across five different initial configurations (generated as outlined in Sec.\ \ref{sec:2.2}) for different combinations of target shape indexes of active and passive cells. The transition from the initial (red bars) to the steady-state distribution (blue bars) illustrates how cells reorganise. In agreement with the previous studies \citep{farhadifar2007influence,staple2010mechanics,sandersius2011correlating}, regardless of the values of $p_0^\text{a}$ and $p_0^\text{p}$, the distribution consistently exhibits a pattern where hexagons dominate the tiling, followed by pentagons, heptagons, and quadrilaterals. Furthermore, comparing the initial configuration with the final dynamic configuration reveals a decrease in the fraction of hexagons and an increase in the fraction of pentagons, irrespective of the target shape indexes. Additionally, the increase in the fraction of pentagons at the expense of hexagons is accompanied by the broadening of the distribution to maintain confluence. Lastly, the final distribution of cell neighbour counts remains qualitatively insensitive to the cell neighbour count distribution of the initial configurations.

\begin{figure}[tbh!]
\centering
  \includegraphics[width=0.66\linewidth]{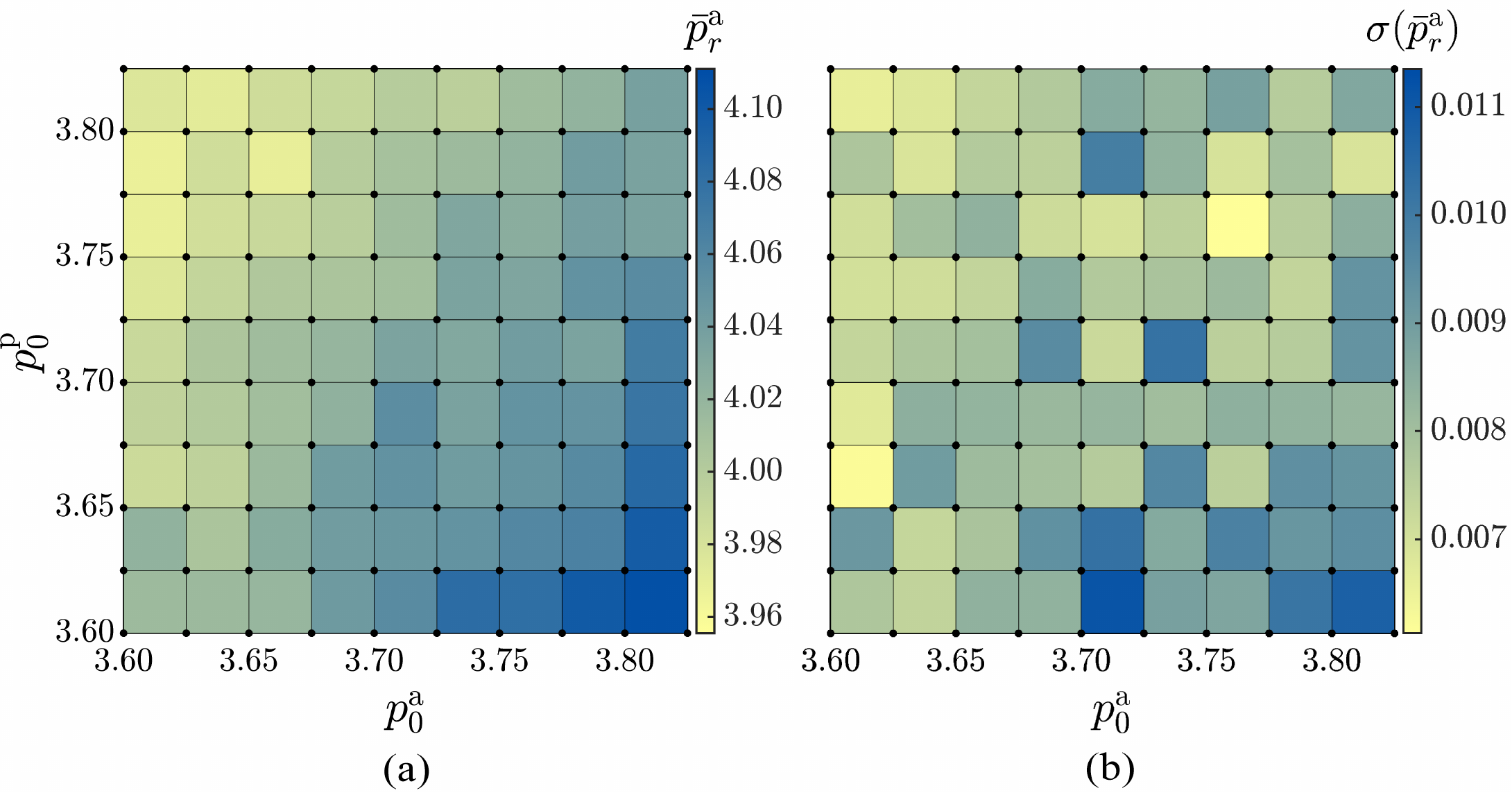}
\caption{\label{fig:state_diag_realized_p0}The dependence of (a) average value of the realised steady-state shape index of active cells, $\bar{p}_\text{r}^\text{a}$, and (b) corresponding standard deviation on the target shape indices of active, $p_0^\text{a}$, and passive, $p_0^\text{p}$, cells. Simulations are performed at grid points indicated by black circles, and the colour within each square represents the average of the values at the four corners of each square.}  
\end{figure}

To gain further insight into the impact of target shape indexes on tissue morphology, we explore the dependence of the mean realised steady-state shape index of the active tissue, $\bar{p}_\text{r}^\text{a}$, and corresponding standard deviation, $\sigma(\bar{p}_\text{r}^\text{a})$, on the input target shape indexes $p_0^\text{a}$ and $p_0^\text{p}$, as shown in Fig.~\ref{fig:state_diag_realized_p0}. $\bar{p}_\text{r}^\text{a}$ is calculated as
\begin{equation}
   \bar{p}_\text{r}^\text{a} = \Big\langle\frac{1}{N^\text{a}(t_i)}\sum_c p_{\text{r},c}^\text{a}(t_i)\Big\rangle_t,
\end{equation}
where $p_{\text{r},c}^\text{a}(t_i) = P_c(t_i)/\sqrt{A_c(t_i)}$ represents the realised shape index of the cell $c$ at simulation step $t_i$, with $P_c(t_i)$ and $A_c(t_i)$, respectively, being the perimeter and area of the cell $c$. Both the mean realised shape index, $\bar{p}_\text{r}^\text{a}$, and the corresponding standard deviation increases with increasing $p_0^\text{a}$ and decreasing $p_0^\text{p}$. This is consistent with Fig.~\ref{fig:time_evol_Nf_sigma}b, which shows that more divisions and ingressions occur at larger values of $p_0^\text{a}$ and lower values of $p_0^\text{p}$, leading to increased disorder in the active tissue.

In summary, the optimal configuration for efficient tissue proliferation, characterised by maximum cell count and reduced disorder (i.e.\ lower realised shape index), involves stiffer proliferating tissue with larger cells enclosed by softer passive tissue with smaller cells.

 \subsection{Effects of confinement - varying the width of the passive tissue}
 \label{sec:conf_effect}
 
\begin{figure}[tbh!]
\centering
  \includegraphics[width=0.95\textwidth]{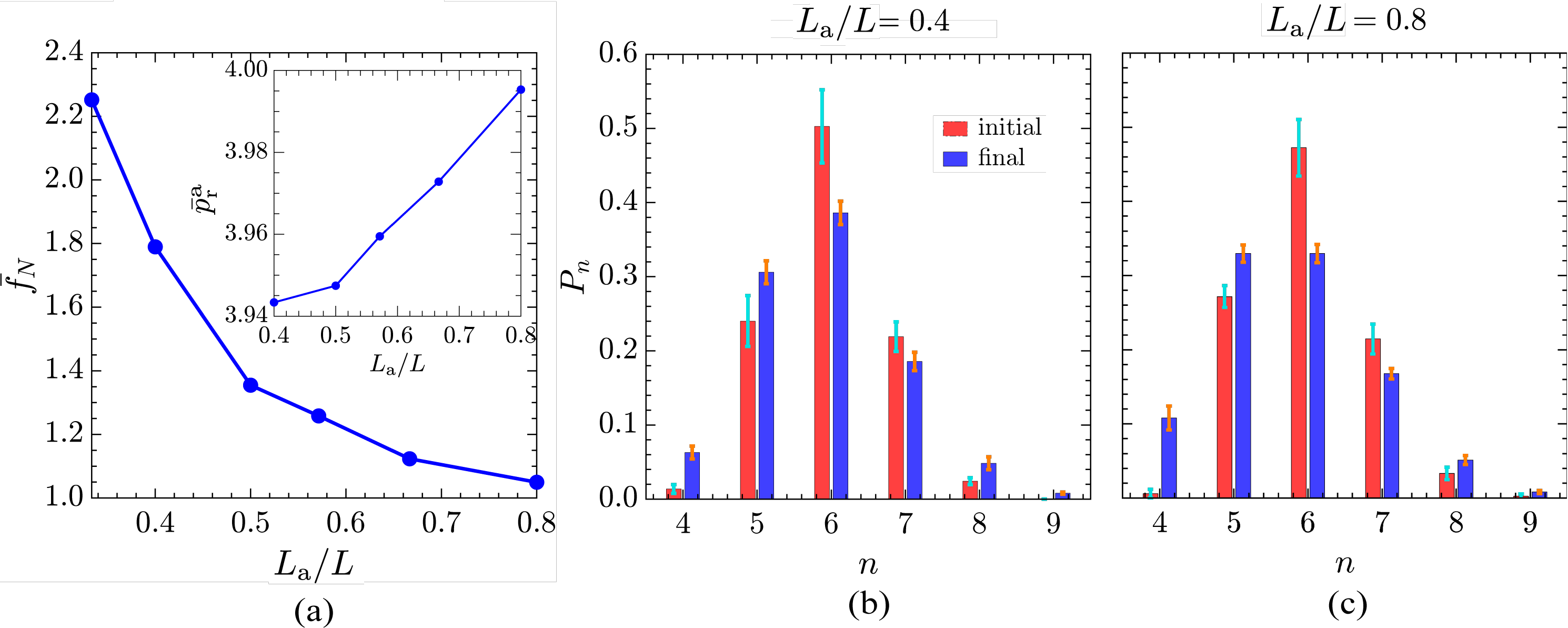}
\caption{\label{fig:effect_of_L} (a) Dependence of the fractional change in the number of active cells, $\bar{f}_N$ [Eqn.\ (\ref{eq:mean_change})], on the width of the passive region, quantified by the ratio $L_\text{a}/L$; Inset: the realised shape index of active cells, $\bar{p}_\text{r}^\text{a}$, as a function of $L_\text{a}/L$, for $p_0^\text{a} = 3.60$ and $p_0^\text{p} = 3.80$. (b) and (c) show the corresponding probability distributions of cell neighbour counts of the active tissue averaged over five different initial configurations for weaker and stronger confinements, respectively. The error bars represent one standard deviation of the five mean values. Here, \(L_\text{a} = 20\) is fixed, and \(L\) is varied to change the ratio \(L_\text{a}/L\).} 
\end{figure}

Finally, we investigate the effect of confinement, quantified by the ratio $L_\text{a}/L$, on the homeostatic state of the active tissue for an optimal configuration of the target shape indices $p_0^\text{a} = 3.60$, $p_0^\text{p} = 3.80$ that result in the maximum number of cells and reduced disorder. By keeping the initial size of the active tissue constant at $L_\text{a} = 20$, we systematically vary $L$ to modify the thickness of the passive tissue and, thus, the strength of the confinement. For each value of $L$, $P_0^\text{p}$ and $ \bar{P}_0^\text{a}$ are calculated as the mean perimeter of the cells in the corresponding initial well-centred Voronoi tiling. Reducing $L_\text{a}/L$ corresponds to weakening the confinement due to the presence of a thick layer of passive tissue between the active tissue and the fixed boundary, which shields the active region from the effects of the fixed boundaries. Conversely, the $L_\text{a}/L\to1$ case corresponds to strong confinement, as there is only a thin layer of passive tissue, and the active patch can easily sense the boundary of the simulation box.

Figure~\ref{fig:effect_of_L} shows how changes in $L_\text{a}/L$ impact the fractional change in the number of active cells and the cell neighbour count distribution. As shown in Fig.~\ref{fig:effect_of_L}a, the cell count monotonically decreases with a decrease in the thickness of the passive tissue (i.e. as confinement strengthens). This is because the clamped boundary emulates a fully rigid tissue, and strengthening the confinement enhances the impact of the fixed boundary on the active tissue. Consequently, making confinement stronger yields effects similar to reducing $p_0^\text{p}$ (Fig.~\ref{fig:nf_af_state_diag}). These findings are consistent with experimental studies that show how mechanical confinement influences cellular behaviour, e.g.\ cells confined in microenvironments~\cite{dike1999geometric,moriarty2018physical,doolin2020mechanosensing}, such as micropatterned substrates, exhibit reduced proliferation rates \cite{yan2011critical,li2023focus}.

Furthermore, Fig.~\ref{fig:effect_of_L}b and Fig.~\ref{fig:effect_of_L}c show the mean distribution of the number of cell neighbours of the active tissue averaged over five different initial configurations for weaker ($L_\text{a}/L = 0.4$) and stronger ($L_\text{a}/L = 0.8$) confinements, respectively. In weaker confinements, hexagons dominate the tiling, followed by pentagons, and then heptagons, whereas in stronger confinements, there is no significant difference between the fraction of hexagons and pentagons (blue bars). However, comparing the initial equilibrium configuration with the final dynamic configuration reveals a decrease in the fraction of hexagons and an increase in the fraction of pentagons irrespective of confinement. Moreover, the fraction of quadrilaterals is higher in strong confinements compared to weak confinements. Thus, strong confinements favour non-hexagonal cells in the active region, whereas weak confinement favours hexagonal cells (Appendix~\ref{appD}). Additionally, as shown in the inset of Fig.~\ref{fig:effect_of_L}a, the realised shape index of active cells ($\bar{p}_\text{r}^\text{a}$) increases as the confinement becomes stronger. Therefore, the disorder in the active tissue increases as confinement becomes stronger, akin to the effect observed when $p_0^\text{p}$ is decreased.

\section{Summary and Conclusions}
\label{sec:conc}
In this paper, we used the two-dimensional vertex model for cell-level modelling of tissue mechanics to investigate homeostasis in confined epithelial tissues consisting of a central region populated with actively proliferating cells surrounded by non-proliferating passive cells. We showed that the system can reach homeostasis which is a dynamic steady state maintained by a balance between cell divisions and ingressions. We characterised the homeostatic state of the tissue in terms of the fractional change in the number of active cells, the area occupied by the active cells, and the distribution of cell shapes. Notably, these parameters demonstrate sensitivity to the mechanical properties of both the active and passive cells, as quantified by their respective target shape indices. Moreover, we observed that the strength of confinement significantly influences the homeostatic state. Our findings illustrate a simple mechanism that regulates the growth of a confined tissue and establishes homeostasis.

The fractional change in the number of active cells and the area occupied by them increases with an increase in the target shape index of the passive cells and decreases with an increase in the target shape index of the active cells. Thus, a softer active tissue with smaller cells (higher target shape index) confined by a stiffer passive tissue with larger cells (lower target shape index) results in controlled growth. However, this configuration results in greater disorder within the proliferating tissue, as indicated by a significantly higher realised shape index. Additionally, as confinement gets stronger, i.e. as the thickness of the passive region decreases, the cell count decreases, while the realised shape index increases, mirroring the effect of decreasing the shape index of the passive cells. However, strong confinement also affects the cells' neighbour count distribution, favouring non-hexagonal cells.

This study emphasizes the central role of cells' mechanical properties, alongside the tissue environment, in regulating homeostasis, providing valuable insights into tissue maintenance. By demonstrating the sensitivity of the homeostatic state to variations in mechanical factors, such as confinement strength and tissue elastic properties, our findings highlight the intricate interplay between mechanical cues and cellular behaviour. These insights could help guide the design of effective tissue scaffolds that replicate the mechanical environment of natural tissues \cite{griffith2006capturing,o2011biomaterials}. Additionally, tissue homeostasis is often disrupted in various diseases, and this work highlights the contribution of mechanical properties of the tissues to pathological conditions \cite{mierke2014fundamental,hanahan2011hallmarks}.

Future work could extend the model to three-dimensional tissues to better capture the complexity and architecture of real tissues \cite{okuda2018combining,rozman2020collective,vzeleznik2024graph}. Additionally, incorporating detailed biochemical signalling pathways will offer a more comprehensive understanding of the interplay between mechanical and biochemical signals in tissue regulation. Including internal dissipation mechanisms would enhance the model’s accuracy in representing embryonic development, especially in environments without a substrate \cite{najera2020cellular,rozman2023dry}. Another promising direction is to explore dynamically varying mechanical properties to mimic the changing physiological conditions seen in living organisms \cite{yan2024bayesian,ogita2022image}. Furthermore, considering the cell cycle~\cite{gradeci2021cell,carpenter2024mechanical} and the impact of the local mechanical environment on cell growth and division would enhance the model's ability to more accurately reflect the behavior of biological tissues. These enhancements would make the model more quantitative, thereby potentially making it applicable to specific biological systems.\\

\newpage
\begin{appendices}
\section*{Appendix A: Active cell count - sensitivity to initial configuration}
\refstepcounter{section}{}\label{appA}

\label{fn_sens}

Here, we discuss the sensitivity of the change in active cell count to the initial configuration. Different initial configurations are generated according to the procedure outlined in Section~\ref{sec:2.2}.

\begin{figure}[tbh!]
\centering
  \includegraphics[width=0.66\textwidth]{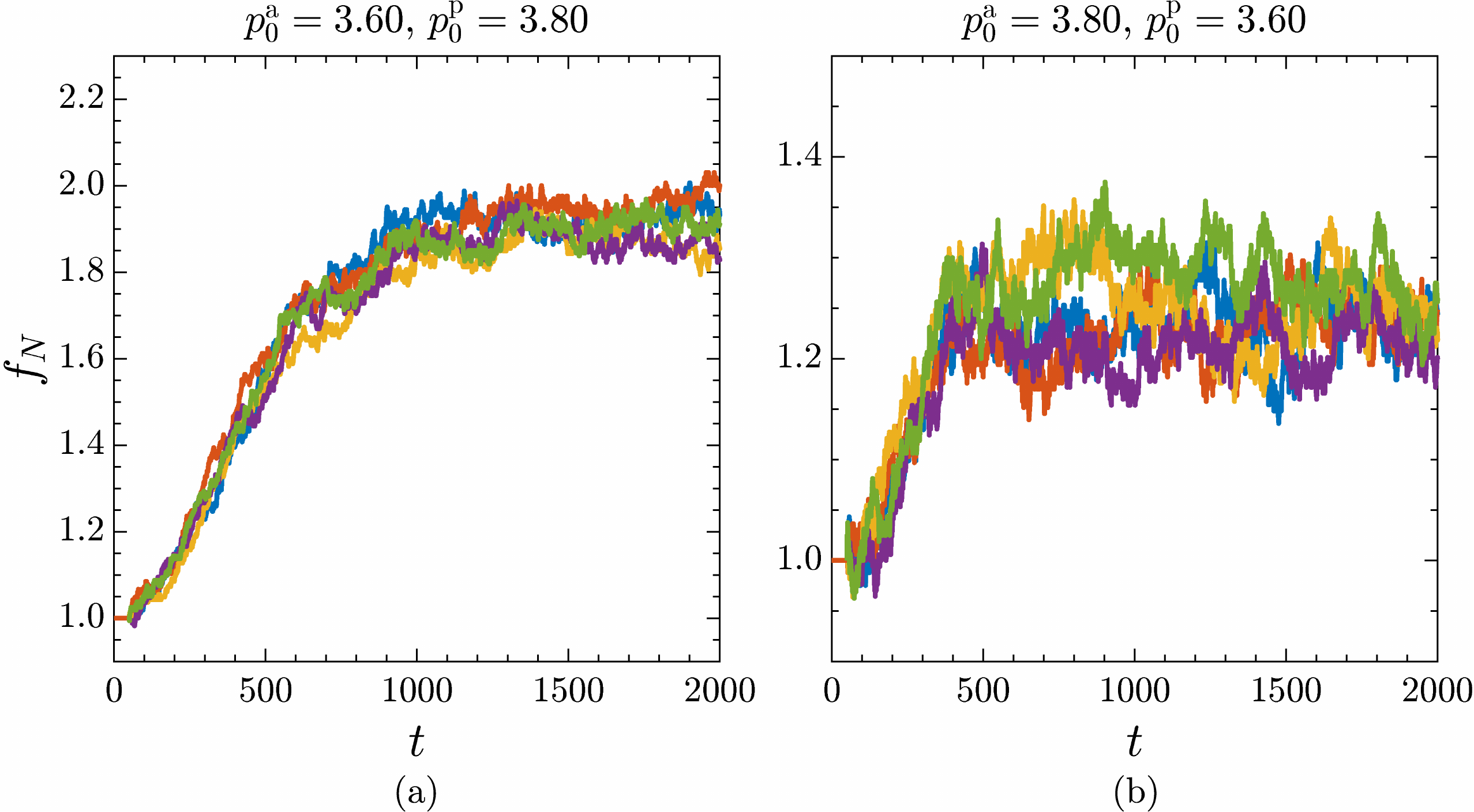}
\caption{\label{fig:vary_reals_fns} The time dependence of the fractional change in the number of active cells is shown for five different initial configurations for (a) $p_0^\text{a} = 3.60$, $p_0^\text{p} = 3.80$, and (b) $p_0^\text{a} = 3.80$, $p_0^\text{p} = 3.60$. Each curve in both panels corresponds to a different initial configuration.}
\end{figure}

Figure~\ref{fig:vary_reals_fns} shows the time evolution of the fractional change in the number of active cells for five different realisations. Specifically, it highlights the shape index combinations resulting in high ($p_0^\text{a} = 3.60$, $p_0^\text{p} = 3.80$) and low ($p_0^\text{a} = 3.80$, $p_0^\text{p} = 3.60$) changes in the active cell count. The final cell count of active cells is nearly insensitive to the initial configuration. Additionally, as discussed in Sec.~\ref{sec:temporal_dynamics}, $f_N$ increases and saturates to a mean value.

\vspace{-0.5cm}
\section*{Appendix B: Analysis for $\alpha = 15$, $\beta = 6$, and $g = 0.006$}
\refstepcounter{section}{}\label{appB}

\begin{figure}[tbh]
  \centering
  \includegraphics[width=0.8\linewidth]{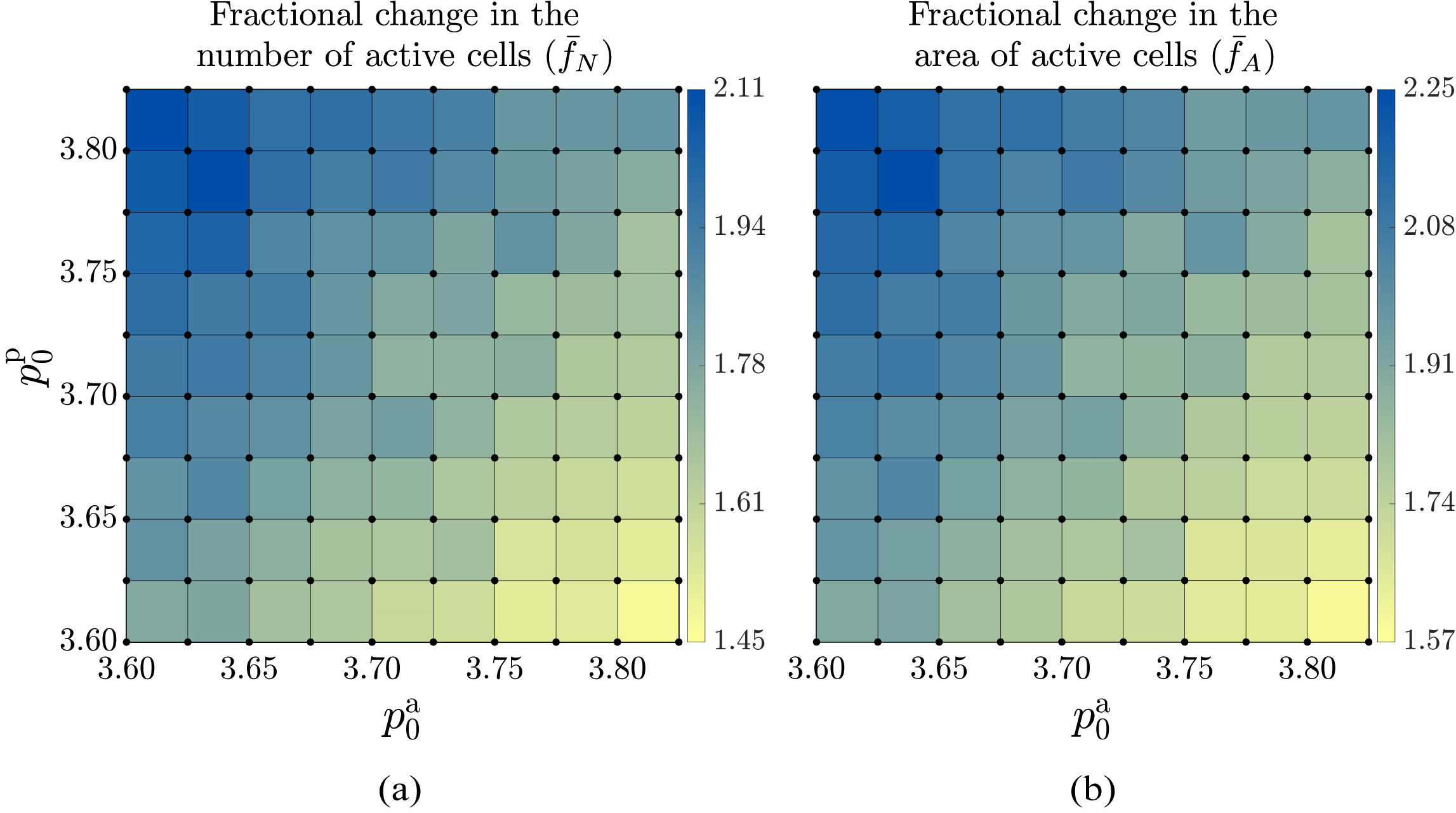}
\caption{\label{fig:fbar_alpha_15} (a) The average number of active cells, $\bar{f}_N$, and (b) the average area occupied by the active cells, $\bar{f}_A$, [Eqns.~(\ref{eq:mean_change})] in the steady state as a function of the target shape index of active ($p_0^\text{a}$) and passive ($p_0^\text{p}$) cells. In panel (b) the data is normalised with respect to the configuration after passive relaxation ($t_\text{ref}=50$). The colour bar applies to both panels. Simulations are performed at grid points indicated by black circles, and the colour within each square represents the average of the values at the four corners of that square. Here, $\alpha = 15$, $\beta = 6$, and $g = 0.006$.}
\end{figure}

In the main text, we focused on the set of parameters where $\bar{f}_N$ and $\bar{f}_A$ are very close to each other, i.e.\ where cells do not significantly change their mean areas from the passive cease. Here, we repeat the analysis in Fig.\ \ref{fig:nf_af_state_diag} but for $\alpha = 15$, $\beta = 6$, and $g = 0.006$ (Fig.\ \ref{fig:fbar_alpha_15}).

\section*{Appendix C: Ratio of the cumulative sum of cell divisions and ingressions}
\refstepcounter{section}{}\label{appC}

\begin{figure}[tbh]
  \centering
  \includegraphics[width=0.45\linewidth]{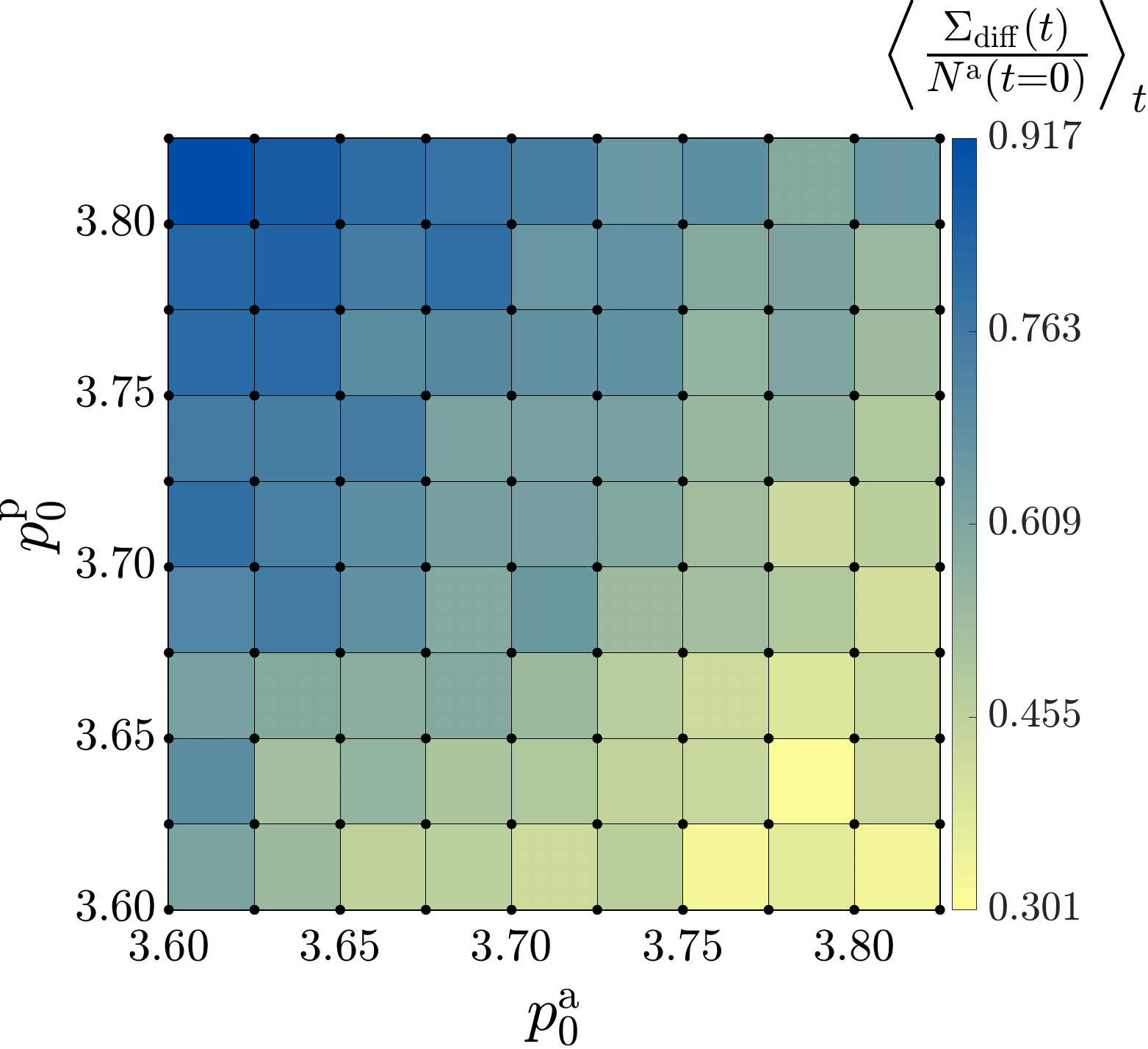}
\caption{\label{fig:cumsum_d_i} The mean value of \textSigma$_\text{diff}(t)$, the difference between the cumulative sum of divisions and ingressions (\textSigma$_d-$\textSigma$_i$), normalized by \(N^\text{a}(t=0)\), calculated over the time interval $\delta t = 500$. Here, $\alpha = 8$, $\beta = 6$, and $g = 0.002$.}
\end{figure}

\section*{Appendix D: Effect of confinement on the polygonal distribution}
\refstepcounter{section}{}\label{appD}

\label{nvsPn}

 \begin{figure}[tbh!]
\centering
  \includegraphics[width=0.33\textwidth]{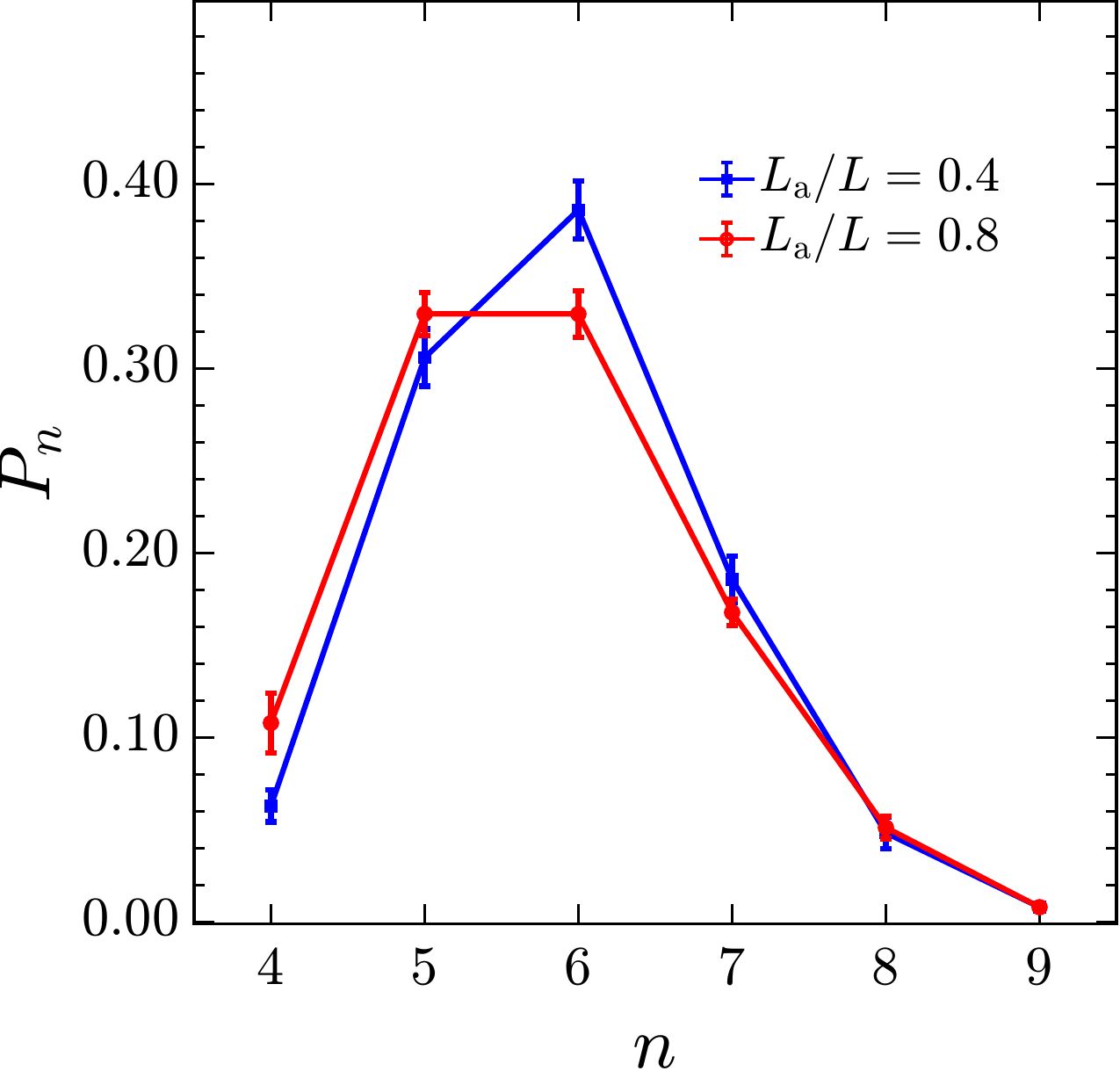}
\caption{\label{fig:nvsPn_lplot}  The steady-state mean probability distributions of cell neighbour counts of the active tissue averaged over five different initial configurations for weaker and stronger confinements. The error bars represent one standard deviation of the five mean values. Here $L_{\text{a}} = 20$ is fixed, and $L$ is varied to change the ratio $L_\text{a}/L$.}
\end{figure}

\end{appendices}

 \newpage   
\begin{acknowledgements}
We wish to thank D.~Barton, E.~Mackay, S.~Terry,  C.~J.~Weijer, G.~Charras, and J.~M.~Yeomans for many helpful discussions. C.K.V.S. and R.S. acknowledge support from the UK Engineering and Physical Sciences Research Council (Award EP/W023946/1). J.R. acknowledges support from the UK Engineering and Physical Sciences Research Council  (Award EP/W023849/1). A.K.~acknowledges support by NSF through Princeton University’s Materials Research Science and Engineering Center DMR-2011750 and by the Project X Innovation Research Grant from the Princeton School of Engineering and Applied Science. This collaboration was initiated during the KITP program ``Symmetry, Thermodynamics and Topology in Active Matter'' (ACTIVE20), and it is supported in part by the National Science Foundation under Grant No.\ NSF PHY-1748958.
\end{acknowledgements}

\section*{Conflict of interest}

The authors declare that they have no conflict of interest.

\bibliographystyle{spphys}       

\begin{thebibliography}{10}
\providecommand{\url}[1]{{#1}}
\providecommand{\urlprefix}{URL }
\expandafter\ifx\csname urlstyle\endcsname\relax
  \providecommand{\doi}[1]{DOI \discretionary{}{}{}#1}\else
  \providecommand{\doi}{DOI \discretionary{}{}{}\begingroup \urlstyle{rm}\Url}\fi

\bibitem{alberts2017molecular}
B.~Alberts, \emph{Molecular Biology of the Cell} (Garland science, 2017)

\bibitem{wolpert2015principles}
L.~Wolpert, C.~Tickle, A.M. Arias, \emph{Principles of development} (Oxford University Press, USA, 2015)

\bibitem{landen2016transition}
N.X. Land{\'e}n, D.~Li, M.~St{\aa}hle, Cell. Mol. Life Sci. \textbf{73}, 3861 (2016)

\bibitem{rodrigues2019wound}
M.~Rodrigues, N.~Kosaric, C.A. Bonham, G.C. Gurtner, Physiol. Rev. \textbf{99}(1), 665 (2019)

\bibitem{burclaff2020proliferation}
J.~Burclaff, S.G. Willet, J.B. S{\'a}enz, J.C. Mills, Gastroenterology \textbf{158}(3), 598 (2020)

\bibitem{weinberg2013biology}
R.A. Weinberg, \emph{The Biology of Cancer} (Garland Science, 2013)

\bibitem{libby2021changing}
P.~Libby, Nature \textbf{592}(7855), 524 (2021)

\bibitem{firestein2003evolving}
G.S. Firestein, Nature \textbf{423}(6937), 356 (2003)

\bibitem{duronio2013signaling}
R.J. Duronio, Y.~Xiong, Cold Spring Harbor Perspect. Biol. \textbf{5}(3), a008904 (2013)

\bibitem{elmore2007apoptosis}
S.~Elmore, Toxicol. Pathol. \textbf{35}(4), 495 (2007)

\bibitem{du2023tuning}
H.~Du, J.M. Bartleson, S.~Butenko, V.~Alonso, W.F. Liu, D.A. Winer, M.J. Butte, Nat. Rev. Immunol. \textbf{23}(3), 174 (2023)

\bibitem{vogel2006local}
V.~Vogel, M.~Sheetz, Nat. Rev. Mol. Cell Biol. \textbf{7}(4), 265 (2006)

\bibitem{stoker1967density}
M.~Stoker, H.~Rubin, Nature \textbf{215}(5097), 171 (1967)

\bibitem{puliafito2012collective}
A.~Puliafito, L.~Hufnagel, P.~Neveu, S.~Streichan, A.~Sigal, D.K. Fygenson, B.I. Shraiman, Proc. Nat. Acad. Sci. \textbf{109}(3), 739 (2012)

\bibitem{lange2024minimal}
S.~Lange, J.~Schmied, P.~Willam, A.~Voss-B{\"o}hme, arXiv preprint arXiv:2403.07612  (2024)

\bibitem{gallaher2019impact}
J.A. Gallaher, J.S. Brown, A.R. Anderson, Sci. Rep. \textbf{9}(1), 2425 (2019)

\bibitem{manzo2020defined}
G.~Manzo, Front. Cell Dev. Biol. \textbf{8}, 804 (2020)

\bibitem{mierke2014fundamental}
C.T. Mierke, Rep. Prog. Phys. \textbf{77}(7), 076602 (2014)

\bibitem{najera2020cellular}
G.S. N{\'a}jera, C.J. Weijer, Mech. Dev. \textbf{163}, 103624 (2020)

\bibitem{najera2023evolution}
G.S. N{\'a}jera, C.J. Weijer, Development \textbf{150}(7) (2023)

\bibitem{asai2023coupling}
R.~Asai, V.N. Prakash, S.~Sinha, M.~Prakash, T.~Mikawa, bioRxiv pp. 2023--05 (2023)

\bibitem{downie1976mechanism}
J.~Downie, Development \textbf{35}(3), 559 (1976)

\bibitem{Treffkorn2022review}
S.~Treffkorn, G.~Mayer, R.~Janssen, Phil. Trans. R. Soc. B \textbf{377}(1865), 20210270 (2022)

\bibitem{szabo2006phase}
B.~Szabo, G.~Sz{\"o}ll{\"o}si, B.~G{\"o}nci, Z.~Jur{\'a}nyi, D.~Selmeczi, T.~Vicsek, Phys. Rev. E \textbf{74}(6), 061908 (2006)

\bibitem{sepulveda2013collective}
N.~Sep{\'u}lveda, L.~Petitjean, O.~Cochet, E.~Grasland-Mongrain, P.~Silberzan, V.~Hakim, PLOS Comput. Biol. \textbf{9}(3), e1002944 (2013)

\bibitem{matoz2017cell}
D.~Matoz-Fernandez, K.~Martens, R.~Sknepnek, J.~Barrat, S.~Henkes, Soft Matter \textbf{13}(17), 3205 (2017)

\bibitem{marth2016collective}
W.~Marth, A.~Voigt, Interface Focus \textbf{6}(5), 20160037 (2016)

\bibitem{mueller2019emergence}
R.~Mueller, J.M. Yeomans, A.~Doostmohammadi, Phys. Rev. Lett. \textbf{122}(4), 048004 (2019)

\bibitem{moure2021phase}
A.~Moure, H.~Gomez, Arch. Comput. Methods Eng. \textbf{28}, 311 (2021)

\bibitem{mueller2021phase}
R.~Mueller, A.~Doostmohammadi, arXiv preprint arXiv:2102.05557  (2021)

\bibitem{glazier1993simulation}
J.A. Glazier, F.~Graner, Phys. Rev. E \textbf{47}(3), 2128 (1993)

\bibitem{hirashima2017cellular}
T.~Hirashima, E.G. Rens, R.M. Merks, Dev. Growth Differ. \textbf{59}(5), 329 (2017)

\bibitem{bi2016motility}
D.~Bi, X.~Yang, M.C. Marchetti, M.L. Manning, Phys. Rev. X \textbf{6}(2), 021011 (2016)

\bibitem{barton2017active}
D.L. Barton, S.~Henkes, C.J. Weijer, R.~Sknepnek, PLOS Comput. Biol. \textbf{13}(6), e1005569 (2017)

\bibitem{honda1978description}
H.~Honda, J. Theor. Biol. \textbf{72}(3), 523 (1978)

\bibitem{farhadifar2007influence}
R.~Farhadifar, J.C. R{\"o}per, B.~Aigouy, S.~Eaton, F.~J{\"u}licher, Curr. Biol. \textbf{17}(24), 2095 (2007)

\bibitem{fletcher2013implementing}
A.G. Fletcher, J.M. Osborne, P.K. Maini, D.J. Gavaghan, Prog. Biophys. Mol. Biol. \textbf{113}(2), 299 (2013)

\bibitem{fletcher2014vertex}
A.G. Fletcher, M.~Osterfield, R.E. Baker, S.Y. Shvartsman, Biophys. J. \textbf{106}, 2291 (2014)

\bibitem{schnyder2017collective}
S.K. Schnyder, J.J. Molina, Y.~Tanaka, R.~Yamamoto, Sci. Rep. \textbf{7}(1), 5163 (2017)

\bibitem{kaiyrbekov2023migration}
K.~Kaiyrbekov, K.~Endresen, K.~Sullivan, Z.~Zheng, Y.~Chen, F.~Serra, B.A. Camley, Proc. Nat. Acad. Sci. \textbf{120}(30), e2301197120 (2023)

\bibitem{camley2014polarity}
B.A. Camley, Y.~Zhang, Y.~Zhao, B.~Li, E.~Ben-Jacob, H.~Levine, W.J. Rappel, Proc. Nat. Acad. Sci. \textbf{111}(41), 14770 (2014)

\bibitem{palmieri2015multiple}
B.~Palmieri, Y.~Bresler, D.~Wirtz, M.~Grant, Sci. Rep. \textbf{5}(1), 11745 (2015)

\bibitem{zhang2020active}
G.~Zhang, R.~Mueller, A.~Doostmohammadi, J.M. Yeomans, J. Royal Soc. Inter. \textbf{17}(169), 20200312 (2020)

\bibitem{monfared2023mechanical}
S.~Monfared, G.~Ravichandran, J.~Andrade, A.~Doostmohammadi, Elife \textbf{12}, e82435 (2023)

\bibitem{graner1992simulation}
F.~Graner, J.A. Glazier, Phys. Rev. Lett. \textbf{69}(13), 2013 (1992)

\bibitem{chiang2016glass}
M.~Chiang, D.~Marenduzzo, Europhys. Lett. \textbf{116}(2), 28009 (2016)

\bibitem{durand2021large}
M.~Durand, PLOS Comp. Biol. \textbf{17}(8), e1008576 (2021)

\bibitem{thuroff2019bridging}
F.~Th{\"u}roff, A.~Goychuk, M.~Reiter, E.~Frey, Elife \textbf{8}, e46842 (2019)

\bibitem{carpenter2024physical}
L.C. Carpenter, S.~Banerjee, bioRxiv pp. 2024--10 (2024)

\bibitem{huang2023bridging}
J.~Huang, H.~Levine, D.~Bi, Soft Matter \textbf{19}(48), 9389 (2023)

\bibitem{teomy2018confluent}
E.~Teomy, D.A. Kessler, H.~Levine, Phys. Rev. E \textbf{98}(4), 042418 (2018)

\bibitem{honda1980much}
H.~Honda, G.~Eguchi, J. Theor. Biol. \textbf{84}(3), 575 (1980)

\bibitem{sego2023general}
T.~Sego, T.~Comlekoglu, S.M. Peirce, D.W. Desimone, J.A. Glazier, Sci. Rep. \textbf{13}(1), 17886 (2023)

\bibitem{bi2015density}
D.~Bi, J.~Lopez, J.M. Schwarz, M.L. Manning, Nat. Phys. \textbf{11}(12), 1074 (2015)

\bibitem{alt2017vertex}
S.~Alt, P.~Ganguly, G.~Salbreux, Phil. Trans. R. Soc. B \textbf{372}(1720), 20150520 (2017)

\bibitem{kursawe2015capabilities}
J.~Kursawe, P.A. Brodskiy, J.J. Zartman, R.E. Baker, A.G. Fletcher, PLoS Comput. Biol. \textbf{11}(12), e1004679 (2015)

\bibitem{xu2015changes}
G.K. Xu, Y.~Liu, B.~Li, Soft Matter \textbf{11}(45), 8782 (2015)

\bibitem{lin2017dynamic}
S.Z. Lin, B.~Li, X.Q. Feng, Acta Mech. Sin. \textbf{33}, 250 (2017)

\bibitem{betts2023epithelial}
J.G. Betts, K.A. Young, J.A. Wise, E.~Johnson, B.~Poe, D.H. Kruse, O.~Korol, J.E. Johnson, M.~Womble, P.~DeSaix, \emph{Epithelial tissue} (ECampus Ontario Pressbooks, 2023)

\bibitem{guillot2013mechanics}
C.~Guillot, T.~Lecuit, Science \textbf{340}(6137), 1185 (2013)

\bibitem{bi2014energy}
D.~Bi, J.H. Lopez, J.~Schwarz, M.L. Manning, Soft Matter \textbf{10}(12), 1885 (2014)

\bibitem{sahu2020linear}
P.~Sahu, J.~Kang, G.~Erdemci-Tandogan, M.L. Manning, Soft Matter \textbf{16}(7), 1850 (2020)

\bibitem{merkel2019minimal}
M.~Merkel, K.~Baumgarten, B.P. Tighe, M.L. Manning, Proc. Natl. Acad. Sci. U.S.A. \textbf{116}(14), 6560 (2019)

\bibitem{wang2020anisotropy}
X.~Wang, M.~Merkel, L.B. Sutter, G.~Erdemci-Tandogan, M.L. Manning, K.E. Kasza, Proc. Natl. Acad. Sci. U.S.A. \textbf{117}(24), 13541 (2020)

\bibitem{tong2022linear}
S.~Tong, N.K. Singh, R.~Sknepnek, A.~Ko{\v{s}}mrlj, PLoS Comput. Biol. \textbf{18}(5), e1010135 (2022)

\bibitem{henkes2020dense}
S.~Henkes, K.~Kostanjevec, J.M. Collinson, R.~Sknepnek, E.~Bertin, Nat. Commun. \textbf{11}(1), 1 (2020)

\bibitem{sknepnek2023generating}
R.~Sknepnek, I.~Djafer-Cherif, M.~Chuai, C.~Weijer, S.~Henkes, e{L}ife \textbf{12}, e79862 (2023)

\bibitem{zwanzig2001nonequilibrium}
R.~Zwanzig, \emph{Nonequilibrium Statistical Mechanics} (Oxford university press, 2001)

\bibitem{du1999centroidal}
Q.~Du, V.~Faber, M.~Gunzburger, SIAM Rev. \textbf{41}(4), 637 (1999)

\bibitem{ustinov2022elastic}
K.~Ustinov, R.~Massab{\`o}, Int. J. Solids Struct. \textbf{248}, 111600 (2022)

\bibitem{nagai2001dynamic}
T.~Nagai, H.~Honda, Philosophical Magazine B \textbf{81}(7), 699 (2001)

\bibitem{hertwig1884problem}
O.~Hertwig, \emph{Das Problem der Befruchtung und der Isotropie des Eies: eine Theorie der Vererbung}, vol.~18 (Fischer, 1884)

\bibitem{rozman2023shape}
J.~Rozman, J.M. Yeomans, R.~Sknepnek, Phys. Rev. Lett. \textbf{131}, 228301 (2023)

\bibitem{yamamoto2022non}
T.~Yamamoto, D.M. Sussman, T.~Shibata, M.L. Manning, Soft Matter \textbf{18}(11), 2168 (2022)

\bibitem{lin2023structure}
S.Z. Lin, M.~Merkel, J.F. Rupprecht, Phys. Rev. Lett. \textbf{130}(5), 058202 (2023)

\bibitem{rozman2023dry}
J.~Rozman, K.V.S. Chaithanya, J.M. Yeomans, R.~Sknepnek, arXiv preprint p. arXiv:2312.11756 (2023)

\bibitem{AJM_git}
R.~Sknepnek.
\newblock Active junction model.
\newblock \url{https://github.com/sknepneklab/ActiveJunctionModel} (2022)

\bibitem{kaliman2021mechanical}
S.~Kaliman, M.~Hubert, C.~Wollnik, L.~Nui{\'c}, D.~Vurnek, S.~Gehrer, J.~Lovri{\'c}, D.~Dudziak, F.~Rehfeldt, A.S. Smith, Phys. Rev. X \textbf{11}(3), 031029 (2021)

\bibitem{o2022tissue}
L.E. O'Brien, Ann. Rev. Cell Dev. Biol. \textbf{38}(1), 395 (2022)

\bibitem{eisenhoffer2012crowding}
G.T. Eisenhoffer, P.D. Loftus, M.~Yoshigi, H.~Otsuna, C.B. Chien, P.A. Morcos, J.~Rosenblatt, Nature \textbf{484}(7395), 546 (2012)

\bibitem{humphrey2014mechanotransduction}
J.D. Humphrey, E.R. Dufresne, M.A. Schwartz, Nat. Rev. Mol. Cell Biol. \textbf{15}(12), 802 (2014)

\bibitem{tong2023linear}
S.~Tong, R.~Sknepnek, A.~Ko{\v{s}}mrlj, Phys. Rev. Res. \textbf{5}(1), 013143 (2023)

\bibitem{staple2010mechanics}
D.~Staple, R.~Farhadifar, J.C. R{\"o}per, B.~Aigouy, S.~Eaton, F.~J{\"u}licher, Eur. Phys. J. E \textbf{33}(2), 117 (2010)

\bibitem{sandersius2011correlating}
S.A. Sandersius, M.~Chuai, C.J. Weijer, T.J. Newman, PLoS One \textbf{6}(4), e18081 (2011)

\bibitem{dike1999geometric}
L.E. Dike, C.S. Chen, M.~Mrksich, J.~Tien, G.M. Whitesides, D.E. Ingber, In Vitro Cell. Dev. Biol. Animal \textbf{35}, 441 (1999)

\bibitem{moriarty2018physical}
R.A. Moriarty, K.M. Stroka, Cell Cycle \textbf{17}(19-20), 2360 (2018)

\bibitem{doolin2020mechanosensing}
M.T. Doolin, R.A. Moriarty, K.M. Stroka, Front. Physiol. \textbf{11}, 365 (2020)

\bibitem{yan2011critical}
C.~Yan, J.~Sun, J.~Ding, Biomater. \textbf{32}(16), 3931 (2011)

\bibitem{li2023focus}
Y.~Li, W.~Jiang, X.~Zhou, Y.~Long, Y.~Sun, Y.~Zeng, X.~Yao, Yale J. Biol. Med. \textbf{96}(4), 527 (2023)

\bibitem{griffith2006capturing}
L.G. Griffith, M.A. Swartz, Nat. Rev. Mol. Cell Biol. \textbf{7}(3), 211 (2006)

\bibitem{o2011biomaterials}
F.J. O'brien, Mater. Today \textbf{14}(3), 88 (2011)

\bibitem{hanahan2011hallmarks}
D.~Hanahan, R.A. Weinberg, Cell \textbf{144}(5), 646 (2011)

\bibitem{okuda2018combining}
S.~Okuda, T.~Miura, Y.~Inoue, T.~Adachi, M.~Eiraku, Sci. Rep. \textbf{8}(1), 2386 (2018)

\bibitem{rozman2020collective}
J.~Rozman, M.~Krajnc, P.~Ziherl, Nat. Commun. \textbf{11}(3805), 1 (2020)

\bibitem{vzeleznik2024graph}
U.~{\v{Z}}eleznik, M.~Krajnc, T.~Sarkar, arXiv preprint arXiv:2408.07551  (2024)

\bibitem{yan2024bayesian}
X.~Yan, G.~Ogita, S.~Ishihara, K.~Sugimura, bioRxiv pp. 2024--04 (2024)

\bibitem{ogita2022image}
G.~Ogita, T.~Kondo, K.~Ikawa, T.~Uemura, S.~Ishihara, K.~Sugimura, PLOS Comp. Biol. \textbf{18}(6), e1010209 (2022)

\bibitem{gradeci2021cell}
D.~Gradeci, A.~Bove, G.~Vallardi, A.R. Lowe, S.~Banerjee, G.~Charras, Elife \textbf{10}, e61011 (2021)

\bibitem{carpenter2024mechanical}
L.C. Carpenter, F.~P{\'e}rez-Verdugo, S.~Banerjee, Biophysical Journal \textbf{123}(7), 909 (2024)

\end{thebibliography}
\providecommand{\noopsort}[1]{}\providecommand{\singleletter}[1]{#1}%

\end{document}